\documentclass{aa}  
\usepackage{threeparttable}
\usepackage{natbib}
\usepackage{xcolor}
\usepackage{placeins}
\usepackage{microtype}
\usepackage{hyperref}
\usepackage{graphicx}
\usepackage{txfonts}

\begin{document}

   \title{Constraining nuclear star cluster formation using MUSE-AO observations of the early-type galaxy FCC\,47\thanks{Based on observation collected at the ESO Paranal La Silla Observatory,
Chile, Prog. ID 60.A-9192, PI Fahrion}}

\titlerunning{Constraining nuclear star cluster formation using MUSE data of FCC\,47}

   \author{Katja Fahrion\inst{1}
          \and
          Mariya Lyubenova\inst{1}
          \and
          Glenn van de Ven\inst{2}           
          \and
          Ryan Leaman\inst{3}
          \and
          Michael Hilker\inst{1}
          \and
          Ignacio Mart\'{i}n-Navarro\inst{4}\fnmsep\inst{3}         
           \and 
          Ling Zhu\inst{5}
          \and
          Mayte Alfaro-Cuello\inst{3}
          \and
           Lodovico Coccato\inst{1}
           \and
           Enrico M. Corsini\inst{6}\fnmsep\inst{7}         
          \and
          Jes\'{u}s Falc\'{o}n-Barroso\inst{8}\fnmsep\inst{9}
          \and
          Enrichetta Iodice\inst{10}
           \and           
          Richard M. McDermid\inst{11}
          \and
           Marc Sarzi\inst{12}\fnmsep\inst{13}
          \and
          Tim de Zeeuw\inst{14}\fnmsep\inst{15}
          }

   \institute{European Southern Observatory, Karl Schwarzschild Stra\ss{}e 2, 85748 Garching bei M\"unchen, Germany\\
   				\email{kfahrion@eso.org}
         \and 
             Department of Astrophysics, University of Vienna, T\"urkenschanzstrasse 17, 1180 Wien, Austria
             \and 
             Max-Planck-Institut f\"ur Astronomie, K\"onigstuhl 17, 69117 Heidelberg, Germany
               \and 
               University of California Santa Cruz, 1156 High Street, Santa Cruz, CA 95064, USA
               \and 
               Shanghai Astronomical Observatory, Chinese Academy of Sciences, 80 Nandan Road, Shanghai 200030, China
               \and 
                Dipartimento di Fisica e Astronomia 'G. Galilei', Universit\`a di Padova,  vicolo dell'Osservatorio 3, I-35122 Padova, Italy
			\and 
			INAF--Osservatorio Astronomico di Padova, vicolo dell'Osservatorio 5, I-35122 Padova, Italy
			\and 
			Instituto de Astrof\'isica de Canarias, Calle Via L\'{a}ctea s/n, 38200 La Laguna, Tenerife, Spain
			\and 
			Depto. Astrof\'isica, Universidad de La Laguna, Calle Astrof\'isico Francisco S\'{a}nchez s/n, 38206 La Laguna, Tenerife, Spain
			 \and 
             INAF-Astronomical Observatory of Capodimonte, via Moiariello 16, I-80131, Napoli, Italy
               \and 
             Department of Physics and Astronomy, Macquarie University, North Ryde, NSW 2109, Australia
             \and 
             Armagh Observatory and Planetarium, College Hill, Armagh, BT61 9DG, Northern Ireland, UK
             \and 
             Centre for Astrophysics Research, University of Hertfordshire, College Lane, Hatfield AL10 9AB, UK
             \and 
             Sterrewacht Leiden, Leiden University, Postbus 9513, 2300 RA Leiden, The Netherlands
             \and 
             Max-Planck-Institut f\"ur extraterrestrische Physik, Gie\ss{}enbachstraße 1, 85748 Garching bei M\"unchen, Germany
             }

   \date{}

 
  \abstract
   {Nuclear star clusters (NSCs) are found in at least 70\% of all galaxies, but their formation path is still unclear. In the most common scenarios, NSCs form in-situ from the galaxy's central gas reservoir, through merging of globular clusters (GCs), or through a combination of the two.} 
   {As the scenarios pose different expectations for angular momentum and stellar population properties of the NSC in comparison to the host galaxy and the GC system, it is necessary to characterise the stellar light, NSC and GCs simultaneously. The large NSC (r$_\text{eff} = 66$ pc) and rich GC system of the early-type Fornax cluster galaxy FCC\,47 (NGC\,1336) render this galaxy an ideal laboratory to constrain NSC formation.}
   {Using MUSE science verification data assisted by adaptive optics, we obtained maps for the stellar kinematics and for stellar-population properties of FCC\,47. We extracted the spectra of the central NSC and determined line-of-sight velocities of 24 GCs and metallicities of five.} 
   {FCC\,47 shows two decoupled components (KDCs): a rotating disk and the NSC. Our orbit-based dynamical Schwarzschild model revealed that the NSC is a distinct kinematic feature and it constitutes the peak of metallicity and old ages in the galaxy. The main body consists of two counter-rotating populations and is dominated by a more metal-poor population. The GC system is bimodal with a dominant metal-poor population and the total GC system mass is $\sim 17\%$ of the NSC mass ($\sim 7 \times 10^8 M_\sun$).}
   {The rotation, high metallicity and high mass of the NSC cannot be uniquely explained by GC-inspiral and most likely requires additional, but quickly quenched, in-situ formation.  The presence of two KDCs most probably are evidence of a major merger that has altered the structure of FCC\,47 significantly, indicating the important role of galaxy mergers in forming the complex kinematics in the galaxy-NSC system.}
   \maketitle
%

\section{Introduction}
The nuclei of galaxies are extreme environments located at the bottom of the galactic potential well, where numerous complex astrophysical processes take place and various phenomena occur in such as active galactic nuclei caused by accretion onto supermassive black holes (SMBHs), central star bursts and extreme stellar densities. As the properties of a galactic nucleus are related to properties of the host galaxy (e.g. \citealt{Magorrian1998, KormendyHo2013}), the evolution of nucleus and host seems to be closely linked and thus, understanding the origin of physical properties of the nucleus gives insight into galaxy evolution. Traditionally, the relations between the mass of the SMBH and host galaxy properties were explored, and subsequently, nuclear star clusters (NSCs) were included that extend the observed relations to lower masses of the central massive object (e.g. \citealt{Ferrarese2006, GrahamDriver2007, Georgiev2016, Spengler2017, Ordenes2018}).

NSCs are dense stellar systems that reside in the photometric \citep{Boeker2002a} and kinematic centre \citep{Neumayer2011} of a galaxy, often co-existing with a central black hole \citep[e.g.][]{Seth2008, GrahamSpitler2009, Schodel2009, Neumayer2012, Georgiev2016}. Typical NSCs have sizes similar to globular clusters (GCs, r$_\text{eff} \sim$ 5-10 pc, \citealt{Boeker2002b, Cote2006, Turner2012, denBrok2014, Georgiev2014, Puzia2014}), but exhibit a broader mass range between 10$^5$ and 10$^8 M_\sun$ \citep{Walcher2005, Georgiev2016, Spengler2017}. Their central stellar densities often reach extreme values (up to 10$^5 M_\sun \, \text{pc}^{-3}$, \citealt{Hopkins2009}), only comparable to the densities found in some GCs and ultra compact dwarf galaxies (UCDs, \citealt{Hilker1999b, Drinkwater2000}). Since especially the more massive UCDs are regarded as the nuclear remnants of stripped dwarf galaxies, it is not surprising that massive UCDs and NSCs have similar properties (e.g. \citealt{Phillipps2001, Pfeffer2013, Strader2013, Norris2015}). 

NSCs are very common and can be found in 70\% to 80\% of all galaxies, from dwarf to giant galaxies of all morphologies \citep{Boeker2002a, Cote2006, Georgiev2009, Eigenthaler2018}. \cite{SanchezJanssen2018} found that the nucleation fraction reaches 90\% for galaxies with $M_\ast \approx 10^9 M_\sun$ and declines for both higher and lower masses. NSCs can show complex density distributions with flattening light profiles \citep{Boeker2002a} and they can exhibit multiple stellar populations \citep{Walcher2006, Lyubenova2013, Kacharov2018}, sometimes having a significant contribution from young stellar populations \citep{Rossa2006, Paudel2011}. NSCs in early-type galaxies (ETGs) in the Virgo galaxy cluster were found to cover a broad range of metallicities \citep{Paudel2011} and they are often more metal-rich than their host galaxy \citep{Spengler2017}. In addition, many NSCs have been observed to show significant rotation \citep{Seth2008b, Seth2010, Lyubenova2013, Lyubenova2019} with the NSC of the Milky Way being the most prominent example \citep{Feldmeier2014}. As nuclear stellar disks usually have similar sizes to NSCs \citep{ScorzaBender1995, Pizzella2002}, it is possible that some NSCs are nuclear stellar disks seen face on. 

There are several suggested formation scenarios for NSCs that distinguish between two main pathways: In the in-situ formation scenario, NSCs form directly at the galactic centre from infalling gas \citep{Bekki2006, Bekki2007, Antonini2015}. This formation mode depends on internal feedback mechanisms and the available gas content \citep{MihosHernquist1994, Schinnerer2008}. Different mechanisms for funnelling gas to the centre have been studied, such as magnetorotational instability, gas cloud mergers or instabilities from bar structures \citep{Milosavljevic2004, Bekki2007, Schinnerer2008}. 
On the other hand, NSC formation might happen through gas-free accretion of GCs that have formed at larger galactic radii and migrate inwards due to dynamical friction \citep{Tremaine1975, CapuzzoDolcetta1993, CapuzzoDolcetta2008, Agarwal2011, ArcaSedda2014}. Also, the formation of nuclear stellar disks via this channel has been explored \citep{Portaluri2013}. While the pure in-situ scenario seems to be unable to reproduce some of the observed scaling relations between NSC and host \citep{Antonini2013}, the dry merger scenario has been more successful in this respect \citep{ArcaSedda2014, Hartmann2011}. 

Studies that have compared the predictions from simulations to observational data have shown that most likely both scenarios are realised in nature \citep{Hartmann2011, Antonini2015} and propose that in-situ formation can contribute a large fraction (up to 80\%) to the total NSC mass. Motivated by this, \cite{Guillard2016} proposed a composite "wet migration" scenario where a massive cluster forms in the early stages of galaxy evolution in a gas-rich disk and then migrates to the centre while keeping its initial gas reservoir, possibly followed by mergers with other gas-rich clusters.  

The different formation scenarios impose different expectations on the properties of the NSC compared to the host galaxy and its GC system. In the in-situ scenario, the NSC forms independent of the GC system and one expects to see more rotation in the NSC due to gas accretion from a disk, and higher metallicities as a consequence of efficient star formation and quick metal enrichment (e.g. \citealt{Seth2006}). The NSC can show an extended star formation history as a result from continuing gas accretion and episodic star formation. 
In the dry GC accretion scenario, the NSC should reflect the metallicity of the accreted GCs \citep{Perets2014} and should show less strong rotation due to the random in-fall directions, however, simulations have shown that NSCs formed through GC mergers, in some cases, also can have significant rotation because they share the angular moment of their formation origin \citep{Hartmann2011, Tsatsi2017}. The kinematic and chemical properties of the surviving GC population can provide insight into the potential contributors to the NSC in this dry scenario. However, if the mergers of gas-rich or young star clusters is considered, studies have shown that these can result in a complex star formation history of the NSC \citep{Antonini2014, Guillard2016}. 

Constraining NSC formation observationally is therefore a complex issue and requires a panoramic view of the kinematic and chemical structure of the stellar body, the NSC and GC population of a galaxy. This can be achieved with modern-day wide-field integral field unit (IFU) instruments as they allow a simultaneous observation a galaxy from its very centre to the halo and are able to spatially and spectrally resolve the different components. 

The early-type galaxy FCC\,47 (NGC\,1336) in the outskirts of the Fornax galaxy cluster (r$_\text{proj}$ = 780 kpc to NGC 1399) at a heliocentric distance of 18.3 $\pm$ 0.6 Mpc (10\arcsec $\approx$ 0.9 kpc, \citealt{Blakeslee2009}, see Figure \ref{fig:FCC47_HST_w_contours} and Table \ref{tab:FCC47} for basic information about FCC\,47) offers an ideal laboratory to explore NSC formation. 
Using the \textit{Hubble Space Telescope} (HST) data from the  ACS Fornax Cluster Survey (ACSFCS, \citealt{Jordan2007}), \cite{Turner2012} determined that FCC\,47 has a particularly large NSC with an effective radius of r$_\text{eff} = 0.750 \pm 0.125 \arcsec = 66.5 \pm 11.1 \,\text{pc}$ in the F475W filter ($\approx g$-band, see also the surface brightness profile in Figure \ref{fig:FCC47_HST_w_contours}) and \citet{Jordan2015} identified more than 300 GC candidates associated with FCC\,47. In the recent ACSFCS study of GC specific frequencies of \cite{Liu2019}, FCC\,47 appears as an outlier with a specific frequency $S_{N,z} = 4.61 \pm 0.21$\footnote{$S_{N,z} = N_\text{GC} \times 10^{0.4(M_z + 15)}$ with the total $z$-band magnitude $M_z$ and the number of GCs $N_\text{GC}$} that is much higher than the typical ($\sim 1$) specific frequency of ETGs with similar mass.
The NSC of FCC\,47 was also studied with adaptive optics (AO)-supported high angular resolution integral-field spectroscopy (IFS) observations with the Spectrograph for INtegral Field Observations in the Near Infrared (SINFONI) by \citet{Lyubenova2019} among five other NSCs of ETGs in the Fornax cluster (see also \citealt{Lyubenova2013}). The NSC of FCC\,47 stands out from this sample due to its large size, strong rotation and pronounced velocity dispersion peak.  

In this paper, we aim to constrain the formation of FCC\,47's massive NSC. We use observations with the Multi Unit Spectroscopic Explorer (MUSE) instrument that enables a simultaneous study of the kinematic and chemical properties of FCC\,47's stellar body, its NSC and the GC system. In an accompanying paper \citep{Fahrion2019}, we report the discovery of a UCD in the MUSE field of view (FOV). Based on its size, magnitude and stellar mass, the UCD is consistent with being the remnant NSC of a stripped metal-poor dwarf galaxy with an initial mass of a few 10$^8 M_\sun$, indicating that FCC\,47 might have undergone a minor merger. However, its properties are also consistent with it being the most massive GC of FCC\,47, although its large effective radius of 24 pc makes it an outlier with respect to other GCs.

We describe the MUSE data briefly in Section \ref{sect:data}. We present the methods to extract MUSE spectra in Sect. \ref{sect:spec_extraction} and how kinematic and stellar population properties were extracted in Sect. \ref{sect:fitting}. The results of our kinematic analysis are presented in Sect. \ref{sect:kinematics}, and Section \ref{sect:stellar_pop} gives the results for the stellar population analysis. Section \ref{sect:schwarzschild} describes our Schwarzschild orbit-based model that was used to explore the origin of the dynamical structure of FCC\,47. We discuss our findings in Sect. \ref{sect:discussion} and give our summary and conclusions in Section \ref{sect:conclusion}. 

\begin{table}
    \centering
    \caption{Basic information about FCC\,47.}
    \begin{threeparttable}
    \begin{tabular}{ l  c  c }\hline \hline
    Property & Value & Reference \\ \hline
  RA (J2000)		& 03:26:32.19 & \\
  DEC  (J2000)	&  -35:42:48.80 & \\
 d (Mpc) & 18.3 $\pm$ 0.6 & \cite{Blakeslee2009} \\
 B$_T$ (mag) & 13.34 & \cite{Glass2011} \\
 $r_\text{eff}$ (arcsec) & 30.0 & \cite{Ferguson1989} \\
 $M_\ast$ ($M_\sun$) & 6.4 $\times 10^9$ & \cite{Saulder2016}\tnote{a} \\ \hline
\end{tabular}
\begin{tablenotes}
\item[a] \citet{Liu2019} give $M_\ast = 9.3 \times 10^9 M_\sun$ based on measurements from \cite{Turner2012}.
\end{tablenotes}
    \label{tab:FCC47}
    \end{threeparttable}
\end{table}

\section{MUSE adaptive optics observations}
\label{sect:data}
The MUSE instrument provides the ideal means for our goals of studying the known large NSC and the rich GC system of FCC\,47. MUSE is an integral-field spectrograph mounted at UT4 on the Very Large Telescope (VLT) in Paranal, Chile. Its wide field mode (WFM) provides a continuous FOV of 1\arcmin$\times$1\arcmin\ with a spatial sampling of 0.2\arcsec\,and a spectral resolution of 2.5\,\AA\,(FWHM) at 7000\,\AA\, in an optical wavelength range between 4500 and 9300\,\AA. 

\begin{figure}
    \centering
    \includegraphics[width=0.49\textwidth]{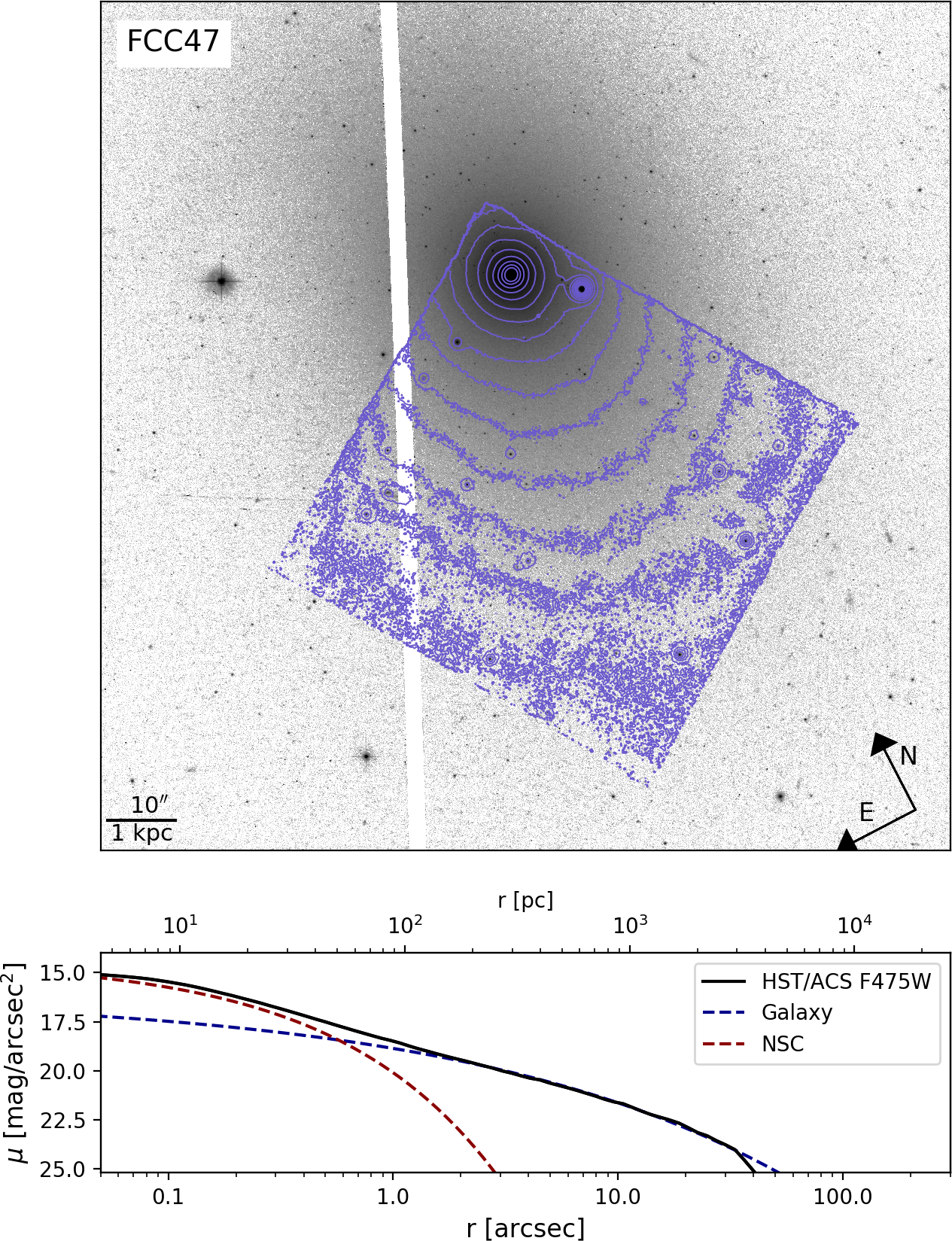}
    \caption{Top: HST image of FCC\,47 in the F475W filter (roughly $g$-band). Surface brightness contours from the collapsed MUSE cube (white light image) are superimposed in purple. The levels are logarithmically spaced, but arbitrarly chosen for illustration purposes. Bottom: Surface brightness profile of FCC\,47 in the F475W filter with a double Sersic profile fit, reproducing the plot from \cite{Turner2012}.}
    \label{fig:FCC47_HST_w_contours}
\end{figure}

Our data were acquired during the Science Verification (SV) phase (60.A-9192, P.I. Fahrion) in September 2017 after the commissioning of the Ground Atmospheric Layer Adaptive Corrector for Spectroscopic Imaging (GALACSI) AO system. We placed the NSC in a corner of the MUSE FOV (see Fig. \ref{fig:FCC47_HST_w_contours}). This off-center positioning was chosen such that a simultaneous observation of the NSC as well as a large number of GCs was possible. 

We designed our observation of FCC\,47 following the observing strategy for the galaxies in the Fornax 3D survey (F3D, \citealt{F3D_Survey}). This survey targets bright galaxies within the virial radius of the Fornax cluster and uses deep MUSE observations to explore the formation and evolution of galaxies in a dense cluster environment. Like for the F3D galaxies, the exposures of FCC\,47 were dithered by a few arcseconds and rotated by 90$^\circ$ to reduce the signature of MUSE's 24 integral-field units on the final image. The data consist of ten exposures with 360 seconds exposure time each. Dedicated sky exposures of three minutes were taken in between the science exposures to perform sky modelling and reduce the contamination from sky emission lines.  Our MUSE + AO data were taken in the nominal mode, meaning a wavelength coverage from 4750 to 9300 \AA. In addition, the regime from 5800 to 6000\,\AA\, is filtered out because the AO facility works with four strong sodium lasers that would saturate the detector in the sodium D line (see Fig. \ref{fig:pPXF_fit}).

Due to unfortunate weather conditions, the seeing during the observation was $\sim$ 1.6\arcsec, however, the AO system was able to reduce this to a final full width at half maximum (FWHM) of the point spread function (PSF) of $\sim 0.7\arcsec$. The data are not ideal to study the effect of the AO system on the MUSE PSF as there is only one bright foreground star in the FOV that is located close to the centre of the galaxy. Its light profile therefore suffers from strong contamination from the bright galaxy background. Nonetheless, we fitted both a Gaussian and Moffat profile using \textsc{imfit} \citep{Erwin2015} and achieved equally good results. We therefore assumed a Gaussian PSF profile with a FWHM of 0.7\arcsec\,needed for the extraction of GC spectra although the precise shape of the PSF is not crucial for our study. The PSF FWHM corresponds to the effective radius of the NSC in FCC\,47 (also 0.7\arcsec), meaning that the especially large NSC allowed us to formally resolve it, while usually NSCs at this distance cannot be resolved with MUSE WFM.

We processed the raw data following the standard reduction pipeline version 2.2 \citep{Weilbacher2014} incoporated in a ESO reflex environment \citep{Freudling2013} that is able to handle the AO data. The data reduction includes bias and overscan subtraction, flat-field correction, wavelength calibration, determination of the line-spread function, and illumination correction. To further reduce the sky residual lines, we applied the Zurich Atmosphere Purge (ZAP) principal component analysis algorithm \citep{Soto2016}.


\section{Extraction of spectra}
\label{sect:spec_extraction}
\subsection{Galaxy stellar light}
The extraction of stellar light kinematics and especially stellar population properties requires a minimum spectral signal-to-noise ratio (S/N). To get a continuous view of these properties from the integrated stellar light, we binned the MUSE data with the Python version of the Voronoi binning routine described in \cite{CappellariCopin2003}. This method enables an adaptive binning to provide a constant spectral S/N per bin. We chose a target S/N of 100\,\AA$^{-1}$ to ensure an accurate extraction of the line-of-sight velocity distribution (LOSVD) parameters as well as the metallicity, stellar age and $\alpha$-element abundance. For the purpose of creating a reliable dynamical Schwarzschild orbit-superposition model, a S/N = 100 is necessary \citep{Krajnovic2015}.
To avoid accreting large bins in the outer regions where errors are strongly non-Poissonian, we excluded all spaxels with S/N $<$ 1 \AA$^{-1}$. The Voronoi-binned MUSE cube contains 435 bins (see Fig. \ref{fig:FCC47_kin}).

\subsection{Globular clusters}
\begin{figure}
    \centering
    \includegraphics[width=0.49\textwidth]{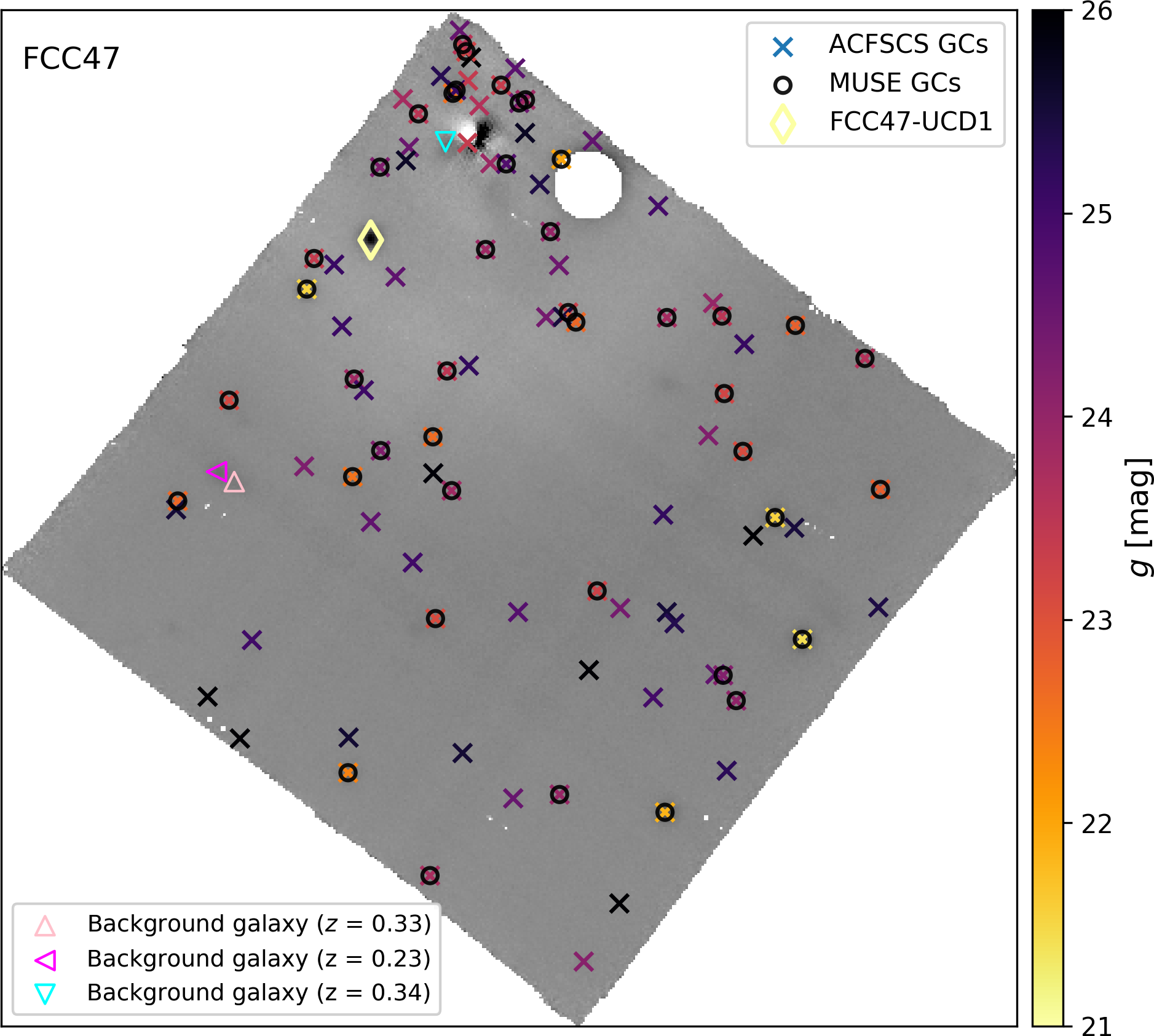}
   \caption{Residual image of FCC\,47 after subtracting a MGE model of the galaxy. The coloured crosses show the position of GC candidates in the HST catalogue of \cite{Jordan2015}, colour-coded by their $g$-band magnitude. The black circles show the position of 42 GC candidates that are found with the \textsc{DAOStarFinder} routine in the MUSE data by cross referencing with the catalogue. The diamond indicates the position of FCC\,47-UCD1 with a $g$-band magnitude of $\sim$ 21 mag. The triangles indicate the positions of emission-line background galaxies we found in the MUSE cube.}
  \label{fig:GC_positions}
  \end{figure}

To extract the spectra of GCs from the MUSE cube, we performed the following steps: First, their spatial location in the unbinned MUSE data was determined. This is difficult in, e.g., the collapsed MUSE image as the majority of GCs (among other point sources) are hidden in the high surface brightness area of the galaxy. For this reason, we determined the individual GC positions after subtracting a model of the galaxy creating a residual image in which the point sources can be identified with standard peak-finding methods. While this has been commonly done in photometric studies, here we had the extra fidelity and luxury of being able to identify the residual signatures of GCs in each single wavelength slice of the MUSE data cube. To ensure a sufficient spatial S/N for the detection of GCs, we created residual images of different slabs from the blue to the red end of the data cube, each containing 50 combined wavelength slices. It is also possible to create a full residual data cube that can be used, e.g., to easily identify astrophysical objects with emission lines, such as planetary nebulae or background star-formation galaxies. In the case of FCC\,47, we used such a residual cube to identify three background galaxies (see Fig. \ref{fig:GC_positions}).

A variety of different approaches were tested to create residual images including a simple unsharp mask approach using a median filter, Multi-Gaussian expansion (MGE, \citealt{Bendinelli1991, Monnet1992, Emsellem1994, Cappellari2002}) modelling and a multi-component \textsc{imfit} model \citep{Erwin2015}. The unsharp mask approach has to be used carefully to avoid losing the faintest GCs and sources in the bright central region. Using MGE or \textsc{imfit} yielded consistent results, but for reasons of time efficiency, we use a simple MGE model using the code described in \cite{Cappellari2002}. However, for other galaxies that have a more complex morphology (e.g. disks with bulges), MGE modelling might be insufficient for removing the galaxy light and \textsc{imfit} should be used in these cases. 

After the MGE model was subtracted from the image, point sources were extracted using \textsc{DAOStarFinder}, a Python module that implements the DAOPHOT algorithm \citep{Stetson1987} into a Python class. This code is usually used to find real point sources like stars, but also detects slightly extended sources very well. \textsc{DAOStarFinder} returns a list of image coordinates of point sources, but these are not exclusively GCs. A contamination from unresolved background galaxies and foreground stars has to be considered. We used the ancillary catalogue of GC candidates from the ACSFCS \citep{Jordan2015} for cross-reference and to build the inital sample of GC candidates. The candidates in this catalog were selected based on their photometric and morphological characteristics, following the approach described in \cite{Jordan2007}. 

We detected 43 GC candidates from the HST catalogue out of 93 in that region and used the associated positional information to extract their spectra from the MUSE cube. We used a PSF-weighted circular aperture extraction assuming a Gaussian PSF with FWHM = 0.7\arcsec\,(3.5 pixels) on the original data cube. Because many of the GC candidates are in high surface brightness areas, we also extracted and subtracted a background spectrum from an annulus aperture around each GC candidate to characterize the local galaxy contribution. The annulus had a width of 5 pixel and the inner radius was placed 8 pixels from the GC centre. The background spectrum was averaged over the used pixels to normalize for the area and no further scaling is applied. For GCs closer than 10\arcsec\,from the galaxy centre we chose an aperture width of 3 pixels and a distance of 6 pixels because of the more strongly varying background. For GCs with distances larger than 30\arcsec, we use annuli with a width of 9 pixels to ensure a high S/N background spectrum. However, for these outer GCs, the background level is low (see surface brightness profile in Fig. \ref{fig:FCC47_HST_w_contours}). The annulus size was chosen to optimize the S/N of the final background-subtracted GC spectrum. Choosing a size that differs by only a few pixels does not affect the recovered radial velocities and mean metallicites considerably, but can increase the uncertainties by a few percent.

Inspecting the GC spectra, we found one GC candidate close in projection to the NSC that is not a GC but rather a star-forming background galaxy at redshift $z \sim$ 0.34 with visible emission lines of hydrogen, nitrogen and oxygen. In addition to this object, we found two other background galaxies with strong emission lines in the MUSE cube as shown in Fig. \ref{fig:GC_positions}. 

\begin{figure}
    \includegraphics[trim={0 0cm 0 0}, clip, width=0.49\textwidth]{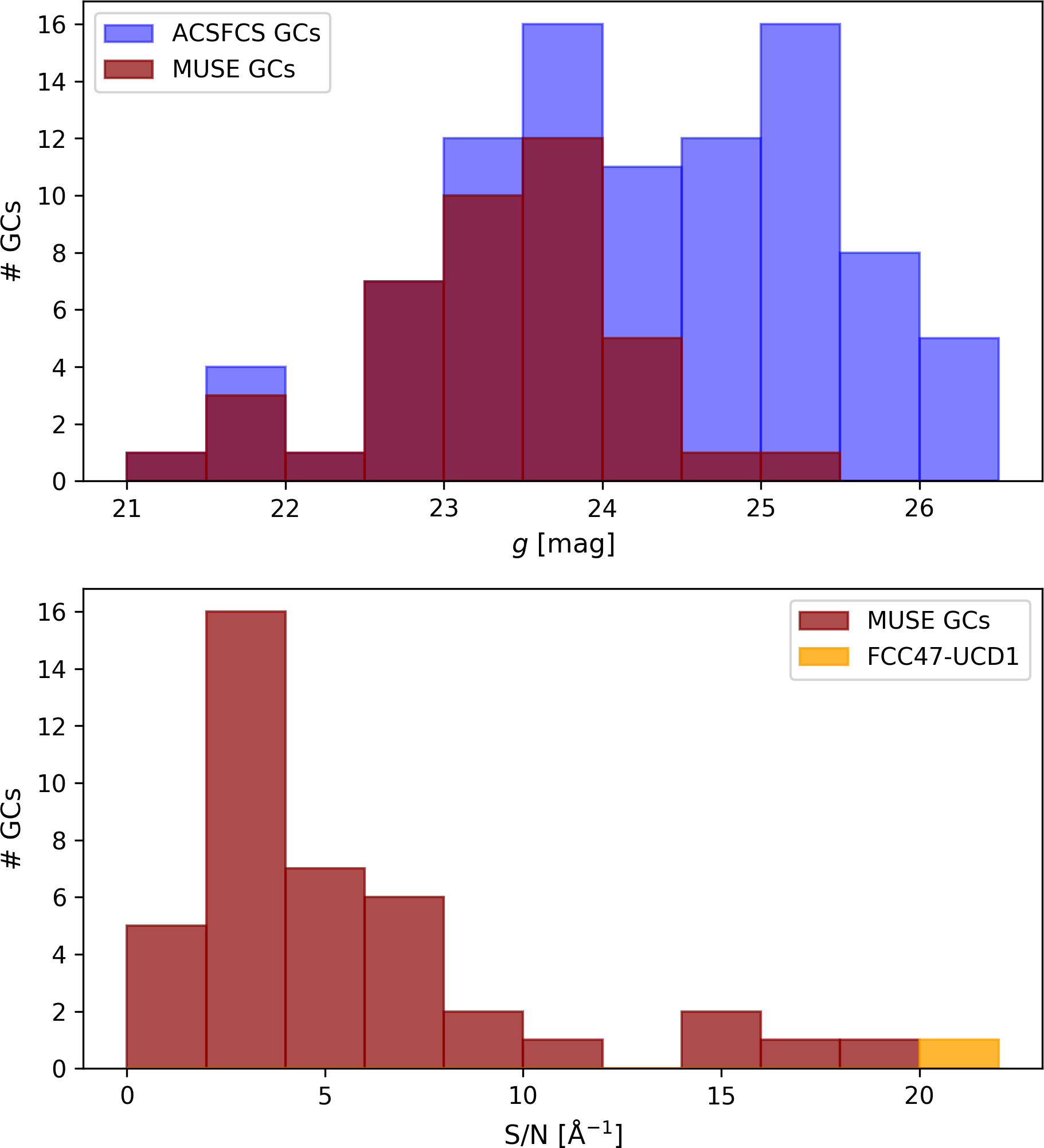} 
    \caption{Top: Distribution of $g$-band magnitudes of the GC candidates in the MUSE FOV from \cite{Jordan2015} (blue). The red distribution highlights the GC candidates found in the MUSE data with our method. Bottom: Histogram of spectral S/N of the GC candidates and FCC\,47-UCD1.} 
    \label{fig:GC_hists}
\end{figure}

The top panel in Fig. \ref{fig:GC_hists} shows a histogram of all GC $g$-band magnitudes from ACSFCS in comparison to the ones that were found in the MUSE data. It reveals that our sample is complete down to a $g$-band magnitude of 23 mag. We are missing five GCs with a magnitude of < 24 mag because they all lie within 10\arcsec\, in projection from the galaxy centre. In this central region, the extraction of GCs is difficult due to the strongly varying galaxy background that is not completely removed by our MGE model, as the residuals show (Figure \ref{fig:GC_positions}). The bottom panel of Fig. \ref{fig:GC_hists} shows the histogram of spectral S/N of the extracted GCs, determined in a continuum region around 6500 \AA. We find 25 GCs with S/N $>$ 3, out of those 17 have a S/N $>$ 5 \AA$^{-1}$ and 5 GCs even reach a S/N $>$ 10 \AA$^{-1}$. A S/N $>$ 3 is required to measure a reliable line-of-sight (LOS) velocity (see Appendix \ref{app:SNR}) and to confirm membership to the FCC\,47 systems. The remaining candidates cannot be confirmed as GCs due to their low S/N. With our approach we found 42 of 93 (45 \%) GC candidates from the ACSFCS catalogue in the MUSE FOV. The detection of GCs with MUSE strongly depends on the PSF FWHM. 

In addition to the GC candidates that were already in the catalogue, it is possible to add sources manually by inspecting their spectra. We reported the discovery of a UCD with a spectral S/N $\sim$ 20 \AA$^{-1}$ in an accompanying paper (\citealt{Fahrion2019}, see also Fig. \ref{fig:GC_positions}). We included FCC-UCD1 in the analysis of the GCs presented below, but highlight it as a distinct object in the associated figures.

\subsection{Spectrum of the nuclear star cluster}
The exceptionally large size of the NSC in FCC\,47 and its brightness allowed us to extract a MUSE spectrum of the NSC. The analysis from \cite{Turner2012} showed that the NSC dominates the inner $\sim$ 0.5\arcsec\,by more than 2 magnitudes. This implies that the inner few spaxels are completely dominated by the light of the NSC, however, there still is a non-negligible contribution from the underlying galaxy. To extract a clean NSC spectrum we used a similar approach as for the GC spectra, i.a. a PSF-weighted circular aperture in combination with a annulus-extracted background spectrum. This approach treats the NSC as another point-source, although it is formally resolved within our MUSE data. Nonetheless, we chose the PSF-weighted extraction to maximize the flux from the NSC. The galaxy background was determined in an annulus aperture that is placed at 5 pixel separation from the central pixel and has a width of 3 pixel. This way, we determined the galaxy contribution directly outside of the NSC, which is crucial to estimate the flux level of the background that shows a strong gradient in the central region. We subtracted the background spectrum from PSF-weighted NSC spectrum both normalised to the same area without additional weighting, i.e., we assume that the background flux level is flat between the centre and the position where we extract the annulus spectrum. This is a justified assumption as e.g. the double S\'{e}rsic decomposition as presented in \cite{Turner2012} shows. From the position where we take the annulus spectrum to the centre, the galaxy component's surface brightness varies by $<$ 1 mag arcsec$^{-2}$, but this depends on the two component S\'{e}rsic fit to the galaxy light profile. For reference, we show the local galaxy spectrum together with the background subtracted NSC spectrum in Fig. \ref{fig:pPXF_fit}. The galaxy background is clearly bluer than the NSC and thus applying an additional weight to it in the subtraction would result in a even redder NSC spectrum. The background-subtracted NSC spectrum has a spectral S/N\,$\sim$\,125\,\AA$^{-1}$.

\section{Extracting kinematics and stellar population properties}
\label{sect:fitting}
\subsection{Fitting with \textsc{pPXF}}
We used the penalized Pixel-fitting (\textsc{pPXF}) method \citep{Cappellari2004, Cappellari2017} to extract the stellar kinematics and population properties, such as mass-weighted mean ages and metallicities, from the MUSE spectra. 
\textsc{pPXF} is a full-spectrum fitting method that uses a penalized maximum likelihood approach to fit a combination of user-provided template spectra to the target spectrum. We chose the single stellar population (SSP) template spectra from the Medium resolution INT Library of Empirical Spectra (MILES, \citealt{Vazdekis2010}). These SSP synthesis models give the spectral energy distributions for stellar populations of a single age and single metallicity. The original MILES spectra have a coverage between 3525 and 7500 \AA. The extended library, called E-MILES, covers a broader wavelength range from 1680 to 50000\,\AA\, including the spectral lines of the Calcium Triplet around 8500\,\AA. The template spectra in both libraries have a spectral resolution of 2.5 \AA\ \citep{FalconBarroso2011}, the mean instrumental resolution of MUSE. We use the description of the line spread function from \cite{Guerou2016} (see also \citealt{Emsellem2019}). Throughout this work, we use BaSTI isochrones \citep{Pietrinferni2004, Pietrinferni2006} and a Milky Way-like, double power law (bimodal), inital mass function (IMF) with a high mass slope of 1.30 \citep{Vazdekis1996}. 

Although the E-MILES templates allow us to fully exploit the broad wavelength coverage of the MUSE instrument, only so-called 'baseFe' models are available. These reproduce the metallicity and light element abundance pattern of the Milky Way and assume $[\text{M}/\text{H}] = [\text{Fe}/\text{H}]$ for high metallicities. For lower metallicities, $\alpha$-enhanced input spectra are used, such that [M/H] is higher than [Fe/H] \citep{Vazdekis2010}. On the other hand, the MILES models have a restricted wavelength range but additionally offer scaled solar models ([$\alpha$/Fe] = 0) and alpha enhanced models ([$\alpha$/Fe] = 0.4 dex). These $\alpha$-variable MILES models allow to study the distribution of $\alpha$-abundances from high S/N spectra. Because the original MILES library only offers two different [$\alpha$/Fe] values, using a regularized fit with \textsc{pPXF} seems unphysical. To create a better sampled grid of SSP models for the fits, we linearly interpolate between the available SSPs to create a grid from [$\alpha$/Fe]  = 0 to [$\alpha$/Fe]  = 0.4 dex with a spacing of 0.1 dex. These models are created under the assumption that the [$\alpha$/Fe] abundances behave linearly in this regime and only give the average [$\alpha$/Fe], however, in reality the abundances of different $\alpha$ elements might be decoupled. The low S/N of the GC spectra make an extraction of [$\alpha$/Fe] abundances challenging. We therefore fitted the GCs with the E-MILES SSP templates to exploit the broad wavelength range of these models, but we noted a metallicity offset of 0.2 dex between E-MILES and scaled solar MILES models. This is caused by the way the E-MILES templates are constructed to match the Milky Way abundance pattern: the input low-metallicity spectra are metal-enhanced \citep{Vazdekis2016}. 

To summarize, we use the baseFe E-MILES templates for the GC spectra on the full MUSE wavelength range to determine estimates of the LOS velocity and metallicities. To extract ages, metallicities and [$\alpha$/Fe] abundances from the high S/N spectra of the NSC and the integrated stellar light, we use the $\alpha$-variable MILES spectra on a wavelength range from 4500 to 7100 \AA.

We did not fit simultaneously for the kinematics and stellar population properties, but used a two-step approach \citep{Sanchez-Blazquez2014}. Each spectrum was first fitted for its LOSVD parameters using additive polynomials of order 25 and no multiplicative polynomials to ensure a well-behaved continuum that is necessary to get an accurate measurement of the LOSVD parameters. In the second step, the LOSVD parameters were fixed for the fit of the stellar population parameters and we used multiplicative polynomials of degree 8 instead of additive polynomials as the relative strength of absorption lines is crucial to determine the stellar population properties. In the end, the \textsc{pPXF} fit returned the LOSVD parameters and the weight coefficients of each input SSP template. In this way, the age and metallicity distributions can be recovered.

\subsection{Galaxy light and NSC spectrum}
\label{sect:spec_gal}
The high S/Ns of the binned stellar light and the NSC spectrum allowed us to accurately measure the LOSVD parameters and recover the distribution of ages and metallicities. We fitted for the LOS velocity, the velocity dispersion and the higher order moments h$_3$ and h$_4$.

We estimated the typical uncertainties of the kinematic fit using a Monte Carlo (MC) approach (e.g. \citealt{Cappellari2004, Pinna2019}). After the first fit, we drew randomly from the \textsc{pPXF} residual in each wavelength bin and added this to the (noise-free) best-fit spectrum to create 300 realisations that were fitted to obtain a well-sampled distribution. The uncertainty is then given by the standard deviation (or 16th and 84th percentile) of this distribution. Testing uncertainties in five different bins, we found maximum uncertainties of 4 km s$^{-1}$ and 6 km s$^{-1}$ for the stellar LOS velocity and velocity dispersion, respectively. For the higher moments of the LOSVD, h$_3$ and h$_4$, we found typical uncertainties of 0.02 km s$^{-1}$ for both.

By construction, using SSP templates discretises the stellar population distribution of a galaxy. Regularisation can be used to find a smooth solution by enforcing smooth variations from one weight to its neighbours.  This is especially crucial for the recovery of star formation histories from the data. Finding the right regularization parameter is, however, non-trivial (see e.g, \citealt{Boecker2019}). Typically, the regularization parameter is determined on a subsample of spectra and then kept fixed (as described in \citealt{McDermid2015}). We calibrated the regularization on a few binned spectra taken from the central part of the galaxy following the recommendation by \cite{Cappellari2017} and \cite{McDermid2015}: Firstly, the noise spectrum is rescaled to obtain a unit $\chi^2$ in an unregularized fit. Then, the regularization parameter is given by the one that increases the $\chi^2$ value of the fit by $\sqrt{2 N_\text{pix}}$, where $N_\text{pix}$ is the number of fitted (unmasked) spectral pixels. This value gives the smoothest solution that is still consistent with the data. Based on this calibration, we used a regularisation parameter of 70 for the binned stellar light fits.

\begin{figure*}
    \centering
    \includegraphics[width=0.99\textwidth]{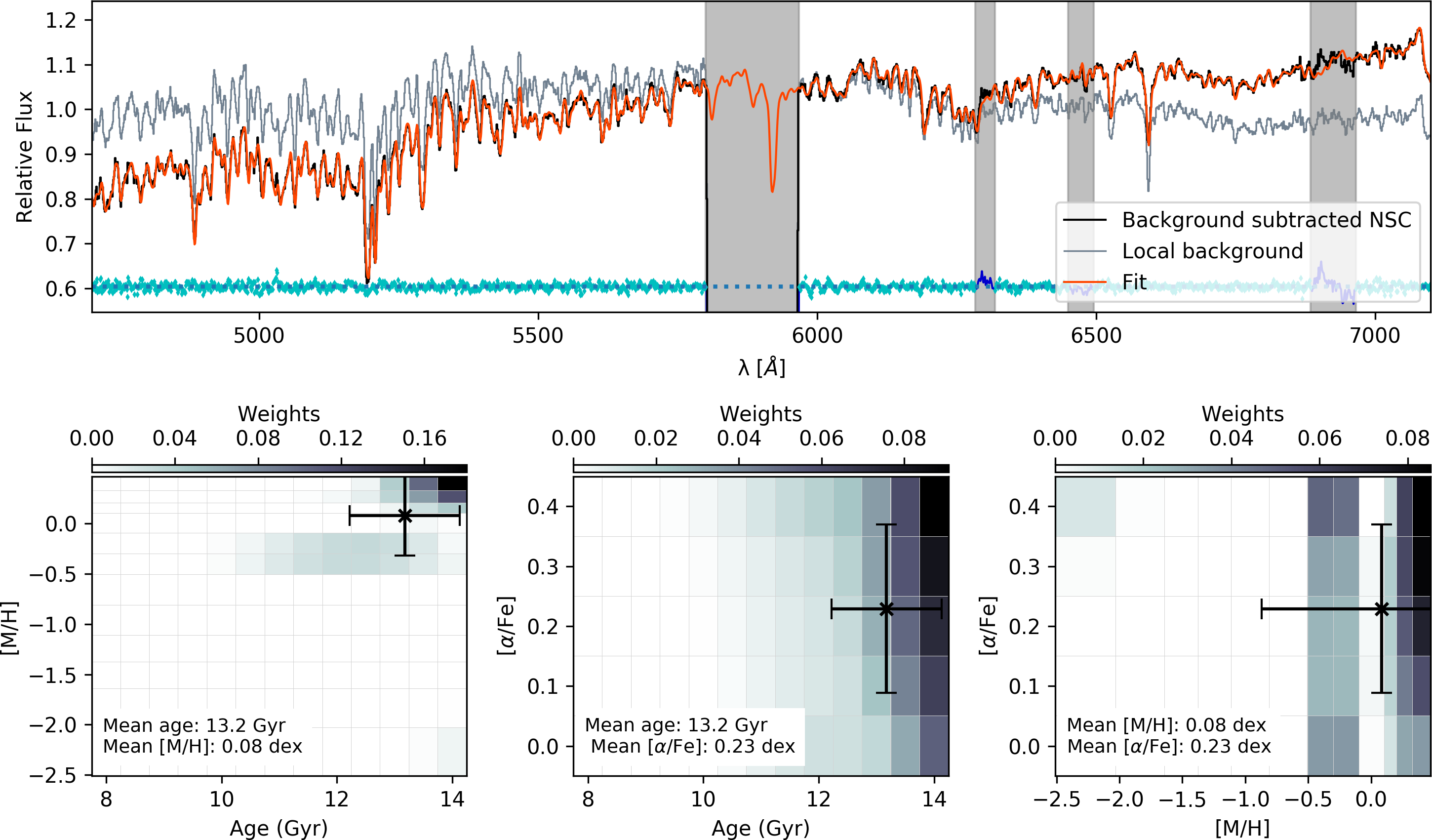}
    \caption{\textsc{pPXF} fit to the background subtracted NSC spectrum. Top: The original normalised spectrum is shown in black, the best-fit spectrum in red. For comparison, the spectrum of the local galaxy background is shown in grey. The residual is shown in blue, shifted to 0.6 for visualisation. Masked regions appear as grey shaded. We masked strong sky line residuals as well as the empty region around the sodium D line that is filtered out due to the AO lasers. Bottom: Weight maps illustrating the age - metallicity (left), age - [$\alpha$/Fe] (middle) and [$\alpha$/Fe] - metallicity (right) grids. The weight of each template used in the fit is given by the greyscale. The black cross shows the weighted mean age, metallicity and $\alpha$-element abundance with errorbars. The weight maps are normalised. A regularisation parameter of 30 was used in this fit.}
    \label{fig:pPXF_fit}
\end{figure*}

We treated the cleaned NSC spectrum similarly to the galaxy light spectra. It has a S/N of 125\,\AA$^{-1}$\, that allows to extract age, metallicities and [$\alpha$/Fe] distributions. Because of this high S/N, we calibrated the regularisation separately and used a regularisation parameter of 30 for the \textsc{pPXF} fit. Figure \ref{fig:pPXF_fit} shows the fit to the cleaned NSC spectrum. The bottom panels of this figure show the grid of age, metallicity and [$\alpha$/Fe] models. The best-fit model is the weighted sum of the SSPs with their weights colour-coded. 

\subsection{Globular clusters}
In contrast to the stellar light, the GCs have a low S/N ($< 20$\,\AA$^{-1}$). This prevented us from detailed stellar population analysis as especially the age is ill-constrained at this spectral quality, and fitting $alpha$-element abundances is even more challenging. In addition, the intrinsic velocity dispersion of typical GCs is $\sim$\,10 km s$^{-1}$, well below the MUSE instrumental resolution and therefore not accessible. Nonetheless, bright GCs allowed us to determine their radial velocity and estimate their mean metallicity. We tested the requirements for the S/N as described in the Appendix  (Section \ref{app:SNR}) and found that radial velocities can be recovered down to a S/N $>$ 3 \AA$^{-1}$, whereas the mean metallicity can be estimated for S/N $>$ 10 \AA$^{-1}$ under the assumption that the GCs are single stellar population objects.
We used the E-MILES SSP templates to fit the GCs, because their broad wavelength range helps to reduce uncertainties.

Since the S/N varies strongly among the GCs, we determined the radial velocity (and, if possible the mean metallicity) with its uncertainty for each GC using 300 realisations in an MC-like fashion as described in Sect. \ref{sect:spec_gal}.
As we could not find indications of young ages ($<$ 10 Gyr) in unrestricted fits, we restricted the SSP library to ages $>$ 10 Gyr to speed up the MC runs for the velocity and metallicity measurements.
 In the best cases, the random uncertainties on the velocity and metallicity, determined from the width of the MC distribution, were $<$ 10 km s$^{-1}$ and $\sim$ 0.10 dex, respectively (see Tab. \ref{tab:GC_properties}).
The afore mentioned regularisation approach can be used for high S/N spectra of the stellar light and the NSC. For the GCs, however, we assumed them to be SSP objects and do not use regularisation in fit. The mean metallicity should be unaffected. As mentioned before, choosing the $\alpha$-enhanced MILES models instead of the E-MILES causes a constant shift of $\sim 0.2$ dex towards higher metallicities, while relative metallicities stay constant. 
\begin{figure*}
\centering
\includegraphics[width=0.99\textwidth]{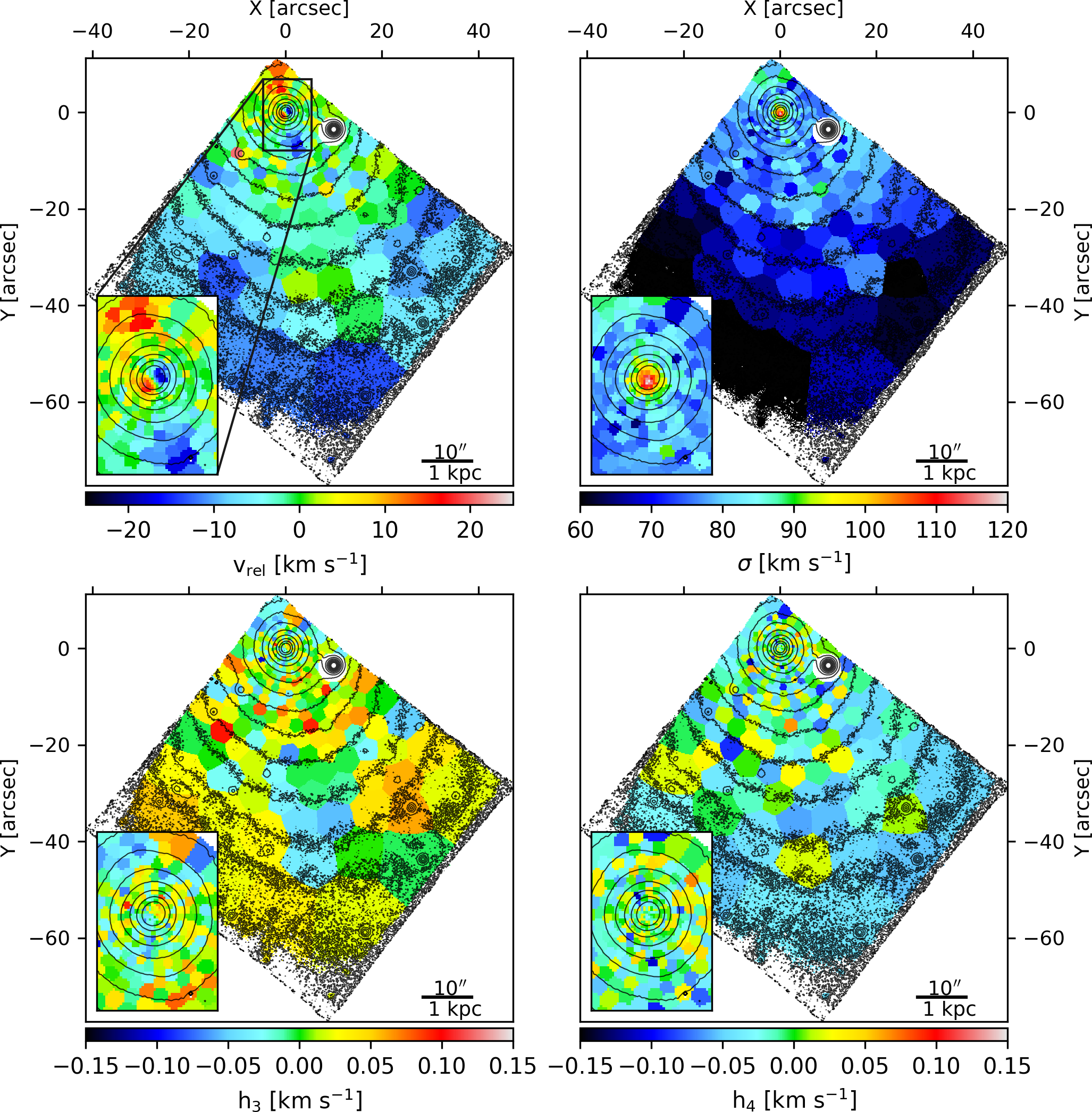}
\caption{Stellar light kinematic maps for FCC\,47 acquired from the Voronoi-binned MUSE cube. Top left: Mean relative velocity of the integrated light with respect to a systemic velocity of 1444 km/s. Top right: Velocity dispersion $\sigma$. Bottom left h$_3$ and bottom right: h$_4$. The inset shows a zoom to the central region to highlight the inner structure. We show the same isophotes as used in Fig. \ref{fig:FCC47_HST_w_contours}. The effective radius of the NSC (0.7\arcsec) corresponds to the radius of the innermost isophote contour.}
\label{fig:FCC47_kin}
\end{figure*}

\section{Kinematics}
\label{sect:kinematics}
\subsection{Galaxy light}
The maps of stellar light LOS mean velocity $v$, velocity dispersion $\sigma$ and the higher order LOSVD parameters h$_3$ and h$_4$ for FCC\,47 are shown in Figure \ref{fig:FCC47_kin}. 
The velocity map revealed FCC\,47's interesting velocity structure. 
While there is no net rotation visible at larger radii in the FOV, a small rotating disk with a diameter of roughly 20\arcsec ($\sim$ 2 kpc) was found near the centre. This feature is rotating with a maximum velocity of  $\lesssim$ 20 km s$^{-1}$, relative to the systemic galaxy velocity of 1444.4 $\pm$ 2.0 $\text{km s}^{-1}$. While this disk structure is rotating along the main photometric axis, the very centre of FCC\,47 is rotating around a different axis offset by $\sim$ 115$^\circ$. The NSC rotation is seen over an extent of $\sim$ 3\arcsec (250 pc, 15 pixels). 

The velocity dispersion map shows a strong peak in the centre, where it reaches values of $\sim$ 125 $\text{km s}^{-1}$ and drops sharply to values of $\sim$ 80 $\text{km s}^{-1}$ before showing a more gradual decrease to values $< 60$ $\text{km s}^{-1}$ in the outskirts. The h$_3$ and h$_4$ maps are rather featureless, but h$_3$ shows anti-correlation to the rotation of the disk structure.

The rotation in the centre matches the rotation of the NSC as studied by \cite{Lyubenova2019} using SINFONI+AO data. In the MUSE data, the NSC does not reach the same maximum velocity and velocity dispersion as in the SINFONI data. The maximum rotation velocity is reduced from $\sim$ 60 km s$^{-1}$ observed with SINFONI down to $\sim$ 20 km s$^{-1}$. Additionally, the velocity dispersion is less peaked in the MUSE data and only reaches 125 km s$^{-1}$ instead of 160 km s$^{-1}$ in the SINFONI data. By convolving the SINFONI maps to match the MUSE PSF with a Gaussian kernel, we found that this difference is caused by the larger MUSE PSF that smears out the NSC's contribution to the velocity structure. This clear link between the SINFONI and MUSE structures shows, however, that the rotating central structure is indeed the NSC of FCC\,47. 

Both the rotating disk structure and the NSC classify as kinematically decoupled components (KDCs) because they are decoupled from the non-rotating main body of FCC\,47.
KDCs were discovered decades ago with long-slit spectroscopy \citep{Efstathiou1982, FranxIllingworth1988}. Large IFS surveys such as SAURON \citep{Bacon2001} or ATLAS3D \citep{Cappellari2011} have revealed that a significant fraction of ETGs have a KDC, especially slow-rotators \citep{Emsellem2007}. Simulations suggest that the formation of KDCs is triggered by major mergers \citep{Jesseit2007, Bois2011, Tsatsi2015}, but once formed, a KDC can be stable over a long time in a triaxial galaxy \citep{vanDenBosch2008, Rantala2019}. KDCs are not rare in slow-rotating galaxies like FCC\,47, however two decoupled components certainly are. 
To explore the nature and origin of FCC\,47's dynamical structure, we constructed an orbit-based dynamical Schwarzschild model (Sect. \ref{sect:schwarzschild}) based on the MUSE kinematic data. 

Using the Wide Field Spectrograph (WiFeS, \citealt{Dopita2010}) on the Australian National University 2.3-m telescope, \citet{Scott2014} presented LOS velocity and velocity dispersion maps of FCC\,47 among nine other ETGs in the Fornax clusters and classified FCC\,47 as a fast rotator based on their data. However, we note that our MUSE maps do not resemble the WiFeS data at all. The reason for this is unknown. 

\subsection{Globular clusters}
\begin{figure}
\centering
\includegraphics[width=0.49\textwidth]{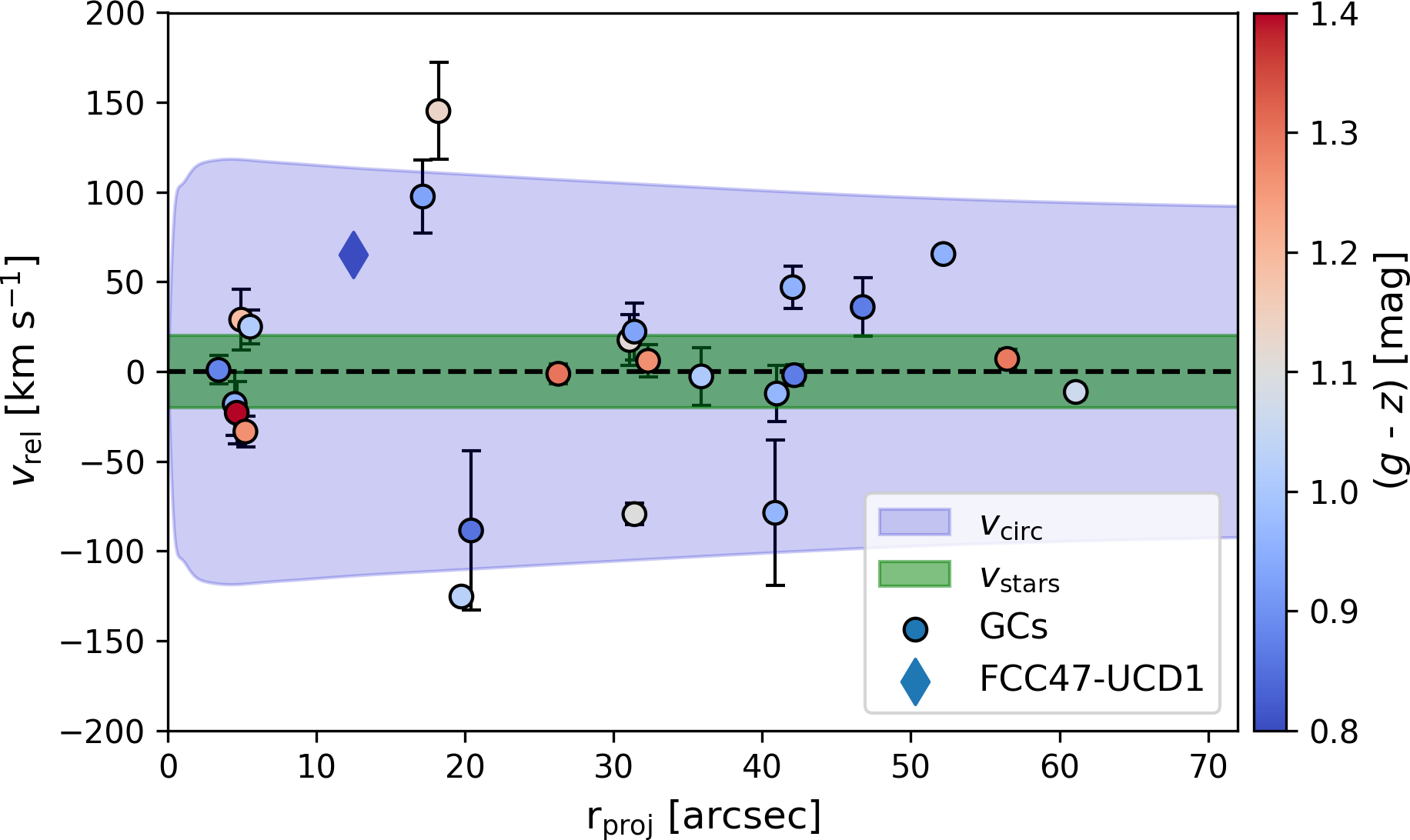}
\caption{Radial profile of LOS velocities of GCs in comparison to the observed scatter in stellar light velocities (green shaded area). The blue shaded area shows the circular velocity profile of FCC\,47 as obtained from the orbit-based dynamical model (Section \ref{sect:schwarzschild}). The colour of the GCs and FCC\,47-UCD1 refers to their ($g - z$) colour in AB magnitudes taken from \cite{Jordan2015}. We show all GCs with S/N $>$ 3 \AA$^{-1}$. The errorbars refer to the 1$\sigma$ random uncertainty from fitting each GC 300 times.}
\label{fig:GC_velocities}
\end{figure}

For the 25 GCs with S/N $>$ 3 \AA$^{-1}$, we determined their radial velocities with \textsc{pPXF} (see Table \ref{tab:GC_properties} in the Appendix). We could not find any signs of rotation in the GC system due to our small sample size.
Figure \ref{fig:GC_velocities} shows the relative GC velocities with respect to the systemic velocity of FCC\,47 as function of their projected distance from the galactic centre, colour-coded by their $(g - z)$ colour from the ACSFCS catalogue \citep{Jordan2015}.
The solid line gives the circular velocity amplitude for FCC\,47 based on the results from our dynamical model (Section \ref{sect:schwarzschild}). FCC\,47-UCD1 is shown as the diamond symbol. The green shaded area indicates the maximum visible rotation amplitude seen in the stellar light map and reveals that the radial velocities for many GCs do not exceed this amplitude, meaning that they are bound to FCC\,47. This is particularly true for the red GC population ($g - z$ > 1.15 mag) as they show less scatter around the systemic velocity of FCC\,47, having a velocity dispersion of 23.0 $\pm$ 3.8 km s$^{-1}$. The blue GCs, however, span a wider range of velocities and have a velocity dispersion of 42.0 $\pm$ 6.8 km s$^{-1}$. The uncertainties refer to the uncertainties from fitting a Gaussian distribution to the histogram of LOS velocities. 

\section{Stellar population properties}
\label{sect:stellar_pop}

\subsection{Galaxy light}
\begin{figure*}
    \centering
    \includegraphics[width=0.99\textwidth]{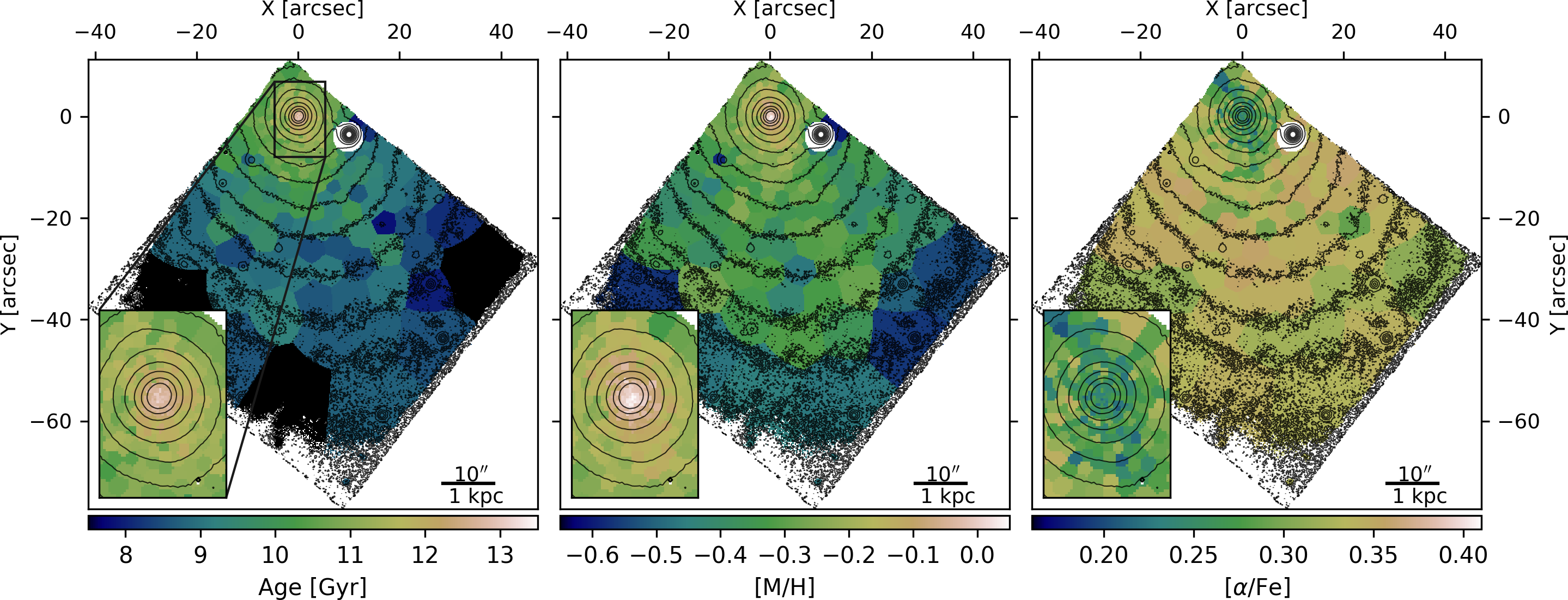}
    \caption{Stellar population maps of FCC\,47. Mean ages (left), metallicities [M/H] (middle) and [$\alpha$/Fe] abundance ratios (right) as determined from a full spectrum fit with \textsc{pPXF} using the alpha-variable MILES template spectra. The maps are mass weighted. We show the same isophotes as in Fig. \ref{fig:FCC47_HST_w_contours}. The inset shows a zoom to the central region and the innermost isophote contour corresponds to the size of the NSC ($R_\text{eff}$ = 0.7\arcsec).}
    \label{fig:stellar_pop_maps}
\end{figure*}

From the weights given by the regularized \textsc{pPXF} fit of the alpha-variable MILES SSP template spectra to the Voronoi binned data, we constructed maps of mean age, metallicity and alpha-element abundance ratios ([$\alpha$/Fe]) as shown in Fig. \ref{fig:stellar_pop_maps}. 
The centre of FCC\,47 reaches super-solar metallicities and shows a strong gradient towards lower metallicities within 10\,\arcsec from the centre. The outskirts reach a metallicity of $\sim$ $-$0.4 dex. The extent of the metal-rich central region is the same as the rotating NSC seen in Figure \ref{fig:FCC47_kin}. As the comparison to SINFONI data showed, this region is completely dominated by the light from the NSC, smeared out by the MUSE PSF. The NSC appears to be significantly metal-richer than the surrounding galaxy. The shown maps were extracted using the alpha-variable MILES models, but we found a very similar metallicity distribution when using the baseFe E-MILES models, however, shifted to lower metallicities by 0.2 dex (see the metallicity gradient shown in Fig. \ref{fig:metal_profile_with_GCs}). This is most likely caused by the rather smooth distribution of alpha-abundances in the outskirts, as the right panel in Fig. \ref{fig:stellar_pop_maps} illustrates. This is in contrast to the MW-like behaviour, encapsulated in the baseFe E-MILES models. The main body of FCC\,47 is alpha-enhanced and shows lower values on the extent of the rotating disk structure. The NSC cannot be seen as a separate component in the [$\alpha$/Fe] map, but shows similar values as the disk of [$\alpha$/Fe] $\sim$ 0.25 dex. 

\begin{figure}
\centering
\includegraphics[width=0.49\textwidth]{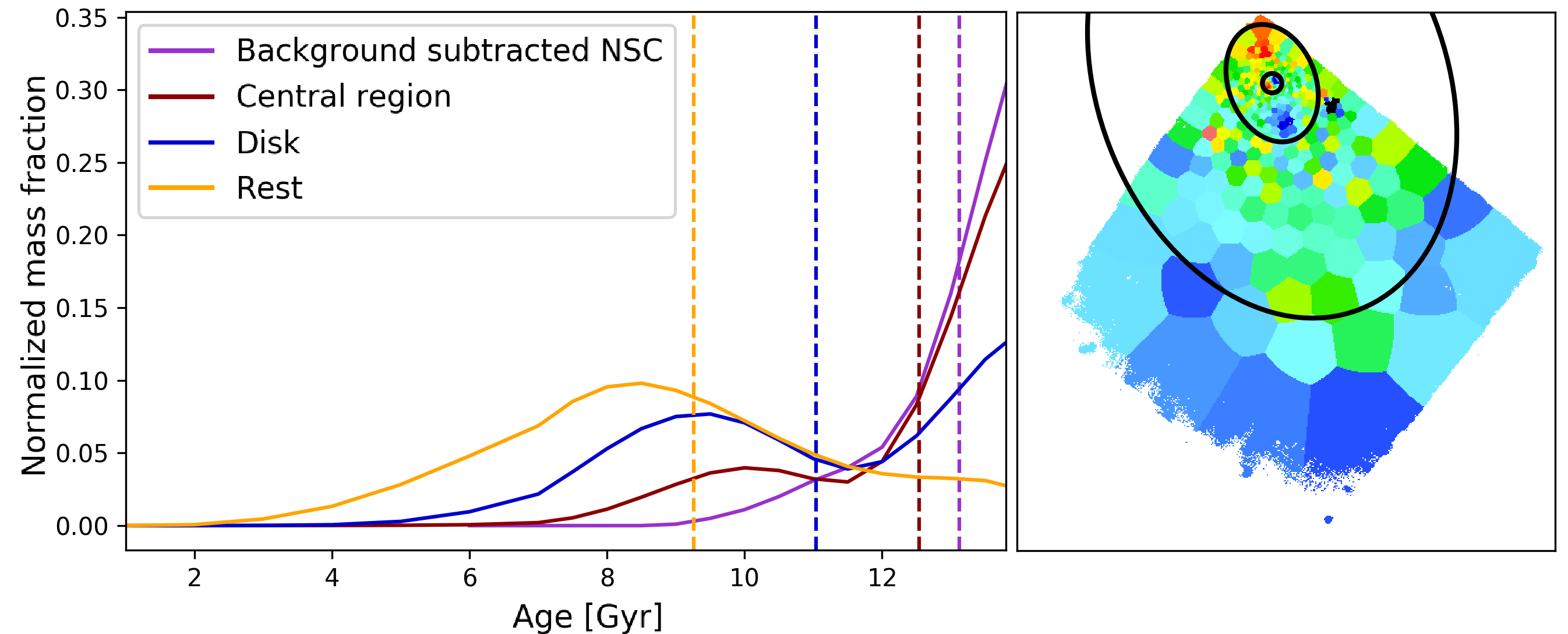}
\caption{Left: Mass weighted, normalised star formation histories of different components of FCC\,47: Background subtracted NSC in purple, central component in red, the disk in blue and the remaining stellar body in yellow. The dotted lines give the mean ages of each component. The normalisation is chosen such that the weights of each component add up to unity. Right: radial velocity map of FCC\,47 as shown in Fig. \ref{fig:FCC47_kin}. The black ellipses illustrate how we divided the different components.}
\label{fig:SFHs}
\end{figure}

FCC\,47 is overall old (> 8 Gyr) with a gradient towards older ages in the centre and the NSC ($\sim$ 13 Gyr). This is also reflected in the star formation histories (SFHs) shown in Fig. \ref{fig:SFHs}. We show the SFH of the background-subtracted NSC, the central region (corresponding to the unsubtracted NSC), the disk structure and the remaining galaxy body. These SFHs were mass-weighted and normalised by their integral. The central component was defined as all bins within 8 pixels (1.6 \arcsec) around the centre. The disk component was described by an ellipse that traces the extent of the rotating disk structure and the galaxy component contains the remaining bins out to a semi major axis distance of 200 pixel (see Fig. \ref{fig:SFHs}). While the background-subtracted NSC is completely dominated by a single star formation peak at very old ages, the disk shows a smaller secondary peak at $\sim$ 8 Gyrs. This secondary peak is visible in the central region, but the comparison to the clean NSC SFH shows that this is due to contamination from the galaxy and not attributed to the NSC itself. The outer galaxy seems to have formed last with the youngest ages found here and this region shows a broad peak in the SFH, centred at $\sim$ 8 Gyr. As the mean ages (dashed vertical lines) illustrate, FCC\,47 has a very old centre and younger outskirts while the disk region shows a transition from old to younger ages and contains two separate populations. 

The mass fraction of the old stellar population ($>$ 11 Gyr) decreases from 93\% in the background-subtracted NSC to 79\% in the centre to 48\% in the disk to 20\% in the rest of the galaxy. Using the same spatial decomposition, we found that the NSC is dominated by a metal-rich population ([M/H] > 0 dex) that constitutes 67\% of the mass in both the cleaned NSC spectrum and the central region. This mass fraction reduces to 47\% in the disk and to 29\% in the outskirts. Comparing the different maps, we find that the metallicity shows a strong gradient of $\sim$ 0.5 dex between NSC and the effective radius of the galaxy (30\,\arcsec) and the mean age decreases from 13 Gyr to 9 Gyr over the same extent, while the [$\alpha$/Fe] abundances show a much milder gradient and increase by $\sim$ 0.1 dex.

\subsection{The nuclear star cluster}
\label{sect:NSC_stellar_pop}
We fitted the background-subtracted NSC spectrum separately to the binned stellar light (see Fig. \ref{fig:pPXF_fit}), allowing a detailed study of the NSC's stellar populations. For comparison, we also show the spectrum of the local background that was used for the subtraction. This background spectrum is clearly bluer than the cleaned NSC spectrum and applying no subtraction would therefore bias the stellar population analysis towards younger ages or more metal-poor populations. From the weight maps shown (Figure \ref{fig:pPXF_fit}), we again found that the NSC shows little substructure in ages and thus must have formed very early on without any later episodes of star formation. However, we found indication of two separate populations that differ in metallicity: There seems to be a dominating metal-rich population and a secondary population with lower metallicities. The dominant metal-rich component constitutes 67\% of the NSC mass. The two populations seem to have similar [$\alpha$/Fe] values. We note, because our models only included [$\alpha$/Fe] values from 0.0 to 0.4 dex, boundary effects from the model grid can introduce uncertainties, but the presence of two populations with different metallicities should be real.

The stellar population analysis allowed us to estimate the stellar mass of the NSC. As we did not have predictions for the mass-to-light ratio from the $\alpha$-variable MILES models, we fitted the NSC again with the E-MILES SSP models and use their predictions for the stellar mass-to-light ratio for each age and metallicity in the HST/ACS F475W filter. The inferred age and metallicity weights were then translated into mass-to-light ratio weights. The weighted mean mass-to-light ratio of the NSC is 5.3 $M_\sun/L_\sun$. Varying [$\alpha$/Fe] will have a small effect on the mass-to-light ratio, however, the uncertainty of the derived mass is governed by the uncertainty on the magnitude. \cite{Turner2012} found an apparent magnitude of the NSC of 16.09 $\pm$ 0.19 mag in the ACS F475W filter ($g$-band), translating to a luminosity of $L= (1.37 \pm 0.23) \times 10^8 L_\sun$, assuming a distance modulus of 31.31 mag \citep{Blakeslee2009}. For the photometric mass, we found $M_\text{NSC, phot} = (7.3 \pm 1.2) \times 10^8 M_\sun$, i.e., $\sim$5 \% of the total stellar mass of FCC\,47. 
The uncertainty was inferred from 1000 draws assuming a Gaussian distribution of the magnitude. 
With this mass, the NSC of FCC\,47 is among the more massive known NSCs as was already expected from its large size. The mass agrees with the enclosed mass profile from the dynamical model (see Figure \ref{fig:schwz_profiles}) within the inner 2.7\arcsec, the extent of the NSC in the MUSE data.

\subsection{Globular clusters}
Measuring the metallicities of the GCs is challenging due to the low S/N. As described in Appendix \ref{app:SNR}, we found that \textsc{pPXF} gives reliable metallicity estimates for GCs with a S/N $>$ 10 \AA$^{-1}$. The MUSE data enabled us to estimate spectroscopic metallicities for a subsample of 5 GCs and FCC\,47-UCD1 (see Tab. \ref{tab:GC_properties}). For this measurement, we used the E-MILES SSP templates because they give the smallest uncertainty due to the broad wavelength range. As mentioned above, using scaled solar MILES templates results in similar metallicities, while the alpha-enhanced E-MILES models gave higher metallicities with a constant offset of 0.2 dex. 

\begin{figure}
    \centering
    \includegraphics[width=0.49\textwidth]{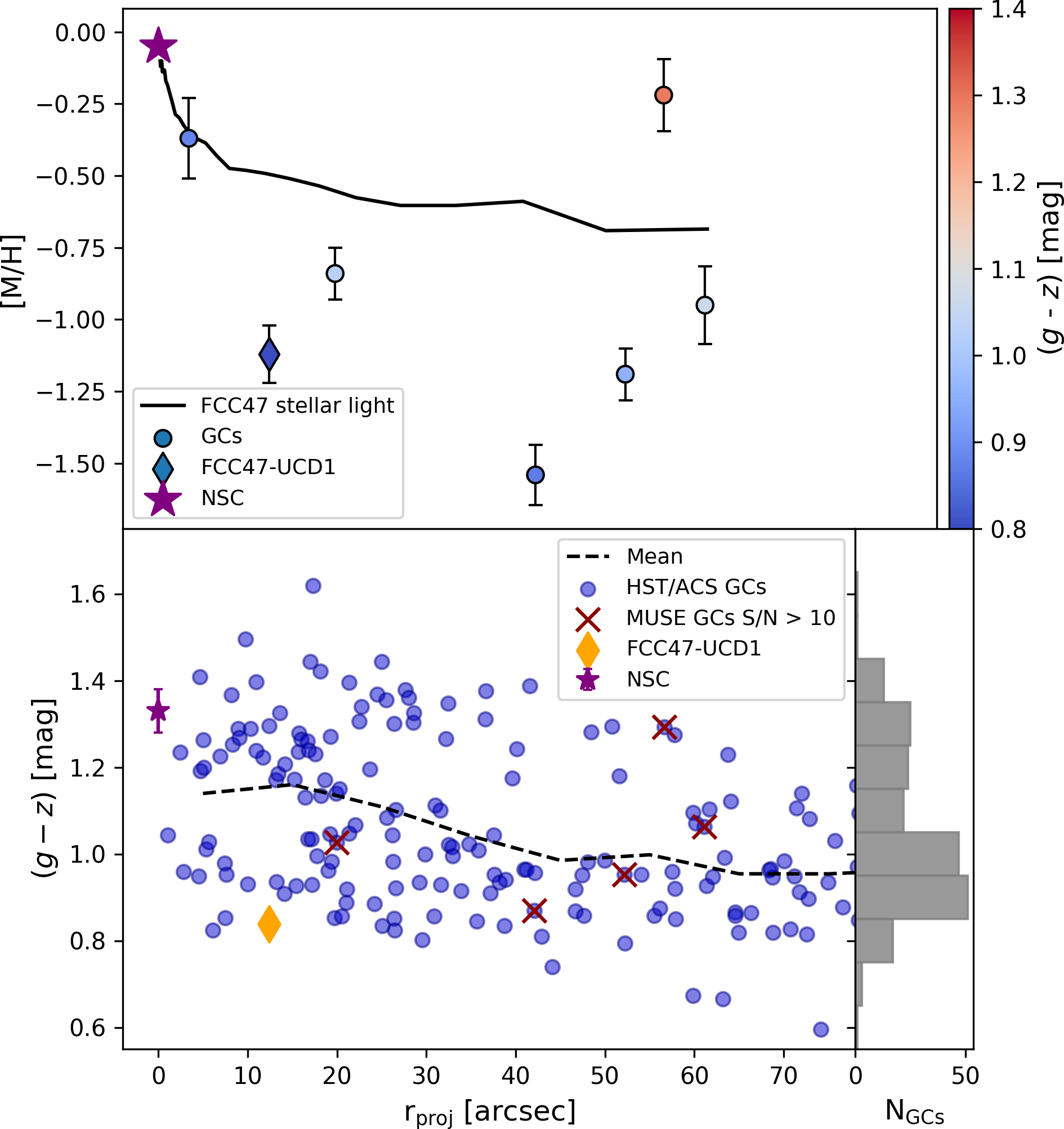}
    \caption{Top: Radial profile of the mean stellar metallicity (black line) with GCs (circles), UCD (diamond) and the NSC (star). The symbols are colour-coded by their ($g - z$) colours from \cite{Jordan2015} and \cite{Turner2012}. Bottom: ($g - z$) colours of the photometric sample of GC candidates from \cite{Jordan2015} versus projected distance. The NSC and FCC\,47-UCD1 are added and we mark the GCs for which we have spectroscopic metallicity measurements. The dotted line is the mean of GC colours in bins of 10 arcsec. The NSC colour is the colour within a 4-pixel aperture on the HST/ACS images \citep{Turner2012}. We also included the histogram of GC colours to illustrate the bimodal colour distribution.}
    \label{fig:metal_profile_with_GCs}
\end{figure}

The metallicities of the five GCs and FCC\,47-UCD1 are shown in a radial profile presented in Fig. \ref{fig:metal_profile_with_GCs}. Four of those GCs have a blue colour and are metal-poor, while the fifth GC is red and metal-rich. Its metallicity is comparable to that of the NSC. Due to the low number of GCs with metallicity estimate, our view on FCC\,47's GC system is biased. We therefore show the radial ($g - z$) colour distribution from the ACSFCS for the GC candidates of FCC\,47 in the bottom panel of Fig. \ref{fig:metal_profile_with_GCs}. We show all GCs that have a probability of being a genuine GC pGC > 0.95. These are 205 GCs. The colours of the NSC \citep{Turner2012} and FCC\,47-UCD1 \citep{Fahrion2019} are added and we mark the GCs with crosses for which we have spectroscopic metallicities. The radial distribution of GC colours also shows the red and blue subpopulations. The red GCs ($g - z$ $>$ 1.15 mag) are more centrally concentrated, which is also reflected in the running mean that shows a radial gradient of mean GC colours from red to blue. 

We used the photometric colours to determine the total mass of the GC system in a similar fashion to the mass of the NSC, however mostly based on photometric SSP predictions that show an age-metallicity degeneracy. Under the assumption that all GCs have an age of 13 Gyr, we used the E-MILES SSP predictions for the HST/ACS filters to convert ($g - z$) colours to metallicities and subsequently to mass-to-light ratios. With the $g$-band magnitudes from the HST catalogue, these are then converted to photometric masses. This way, we estimate the total mass of the GC system (including FCC\,47-UCD1 and all GCs with pGC > 0.5) to be $\sim$ 1.2 $\times 10^8 M_\sun$, 77\,\% of this mass is in the blue GC population ($g - z$ < 1.15 mag), 23\,\% in the red.
This estimate is not precise, but shows that the entire GC system only constitutes $\sim 17\%$ of the mass found in the NSC ($7 \times 10^8 M_\sun$), even though FCC\,47 has a large GC system for its mass. 

In massive ETGs, the mass of the GC system is often found to be proportional to the total stellar mass $M_\text{GC} = (1 - 5) \times 10^{-3} M_\ast$ \citep{SpitlerForbes2009, Georgiev2010, Harris2013}. For FCC\,47, this relation would imply a total GC system mass of $(1.6-8.0) \times 10^7 M_\sun$. While the relation between GC system mass and stellar mass of a galaxy can show significant scatter, it has been established that the total GC system mass forms a near-linear relation with the total halo mass of a galaxy \citep{SpitlerForbes2009, Georgiev2010, Forbes2018}. Using the relation from \citet{SpitlerForbes2009}, the halo mass extrapolation from our dynamical model (Sect. \ref{sect:schwarzschild}) of $1.1 \pm 1.0 \times 10^{12} M_\sun$ would imply a total GC system mass of $\sim 3.5 \times 10^7 M_\sun$.
More recently, \citet{BurkertForbes2019} explored the relation between number of GCs and the halo mass. Using their relation and the number of GCs in FCC\,47 of $N_\text{GC} = 286.6 \pm 13.4$ from \cite{Liu2019}, the expected halo mass would be $\sim 1.4 \times 10^{12} M_\sun$, well within our uncertainties.

\subsection{Metallicity distribution from the centre to the outskirts}
The top panel of Fig. \ref{fig:metal_profile_with_GCs} shows the radial profiles of metallicity for the GCs, the UCD and the NSC in FCC\,47. For this comparison, we exclusively used the metallicites obtained with the baseFe E-MILES SSP templates. The stellar light profile was extracted from the binned map using ellipses with constant position angle of 69$^\circ$ and ellipticity of 0.28. 

This figure shows the variety of different stellar systems that were found within FCC\,47 and their chemical composition. The highest metallicities are found at the centre of the galaxy, more precisely in the NSC that is even more metal-rich than an extrapolation of the galaxy metallicity gradient would imply. This is another indication that the NSC of FCC\,47 is not just the continuation of the stellar light, but instead is a separate component. Most of the GCs and FCC\,47-UCD1, on the other hand, have generally lower metallicities than the underlying stellar body of FCC\,47 and are much more metal-poor than the NSC. However, we found one GC that has a higher metallicity than its local galaxy background, also indicated by its red colour. This GC is only slightly more metal-poor than the NSC and has a projected distance to the centre of $\sim$ 56\arcsec ($\sim$ 50 kpc).

\section{Orbit-based dynamical model}
\label{sect:schwarzschild}
The kinematic maps (Fig. \ref{fig:FCC47_kin}) revealed the unusual and complex dynamical structure of FCC\,47. From the velocity map we could already infer that there are different kinematic components, but a dynamical model is needed to study the orbital structure further. 
In this section we describe the orbit-based dynamical Schwarzschild model of FCC\,47. A detailed description of the concept of Schwarzschild modelling can be found in \cite{vanDenBosch2008, vanDeVen2008} and \cite{Zhu2018a}. Here, we give a short summary.

\subsection{Schwarzschild orbit-superposition technique}
Schwarzschild modelling directly infers the distribution function, that describes the positions and velocities of all stars in a galaxy, as the weights of orbits in the galaxy's gravitational potential.  In this way, not only mass and density profiles can be determined, but it is also possible to explore the orbital structure of a galaxy.

To create a Schwarzschild model, first a suitable model of the gravitational potential has to be constructed. Then, a representative orbital library for this potential is calculated, and finally, a combination of orbits that reproduce the observed light distribution and kinematic maps is found. Each Schwarzschild model depends on a series of parameters that describe the viewing orientation (and hence deprojected shape). The shape parameters $p_\text{min}$ (minimum medium-to-major axis ratio), $q_\text{min}$ (minimum minor-to-major axis ratio) and $u$ (ratio between observed and intrinsic size) can be translated into three viewing angles ($\theta$, $\psi$, $\phi$). Assuming a constant stellar mass-to-light ratio $(M/L)$, the observed light distribution is translated to a 3D mass distribution that constitutes the stellar contribution to the gravitational potential. The dark matter halo is described as a Navarro-Frenk-White halo \citep{NFW1996} with a concentration log($c$) and mass fraction between virial and stellar mass log($M_{\text{vir}}/M_\ast$). Our model also includes a central black hole with a mass of 10$^7 M_\sun$, following a rough estimate with the $M_\ast$-$\sigma$-relation \citep{Ferrarese2000, Gebhardt2000}. We fix this parameter as the resolution of the MUSE data is insufficient to determine the black hole mass directly, so our model has in total six free parameters: three shape parameters (or viewing angles) as well as ($M/L$), $c$ and log($M_{\text{vir}}/M_\ast$). These are iteratively adapted by exploring a user-specified parameter grid. For each model, an orbit library with $\sim$ 10000 orbits is created to fit the observed data and the best model is determined by its minimized $\chi^2$ value.

\begin{figure}
    \centering
    \includegraphics[width=0.5\textwidth]{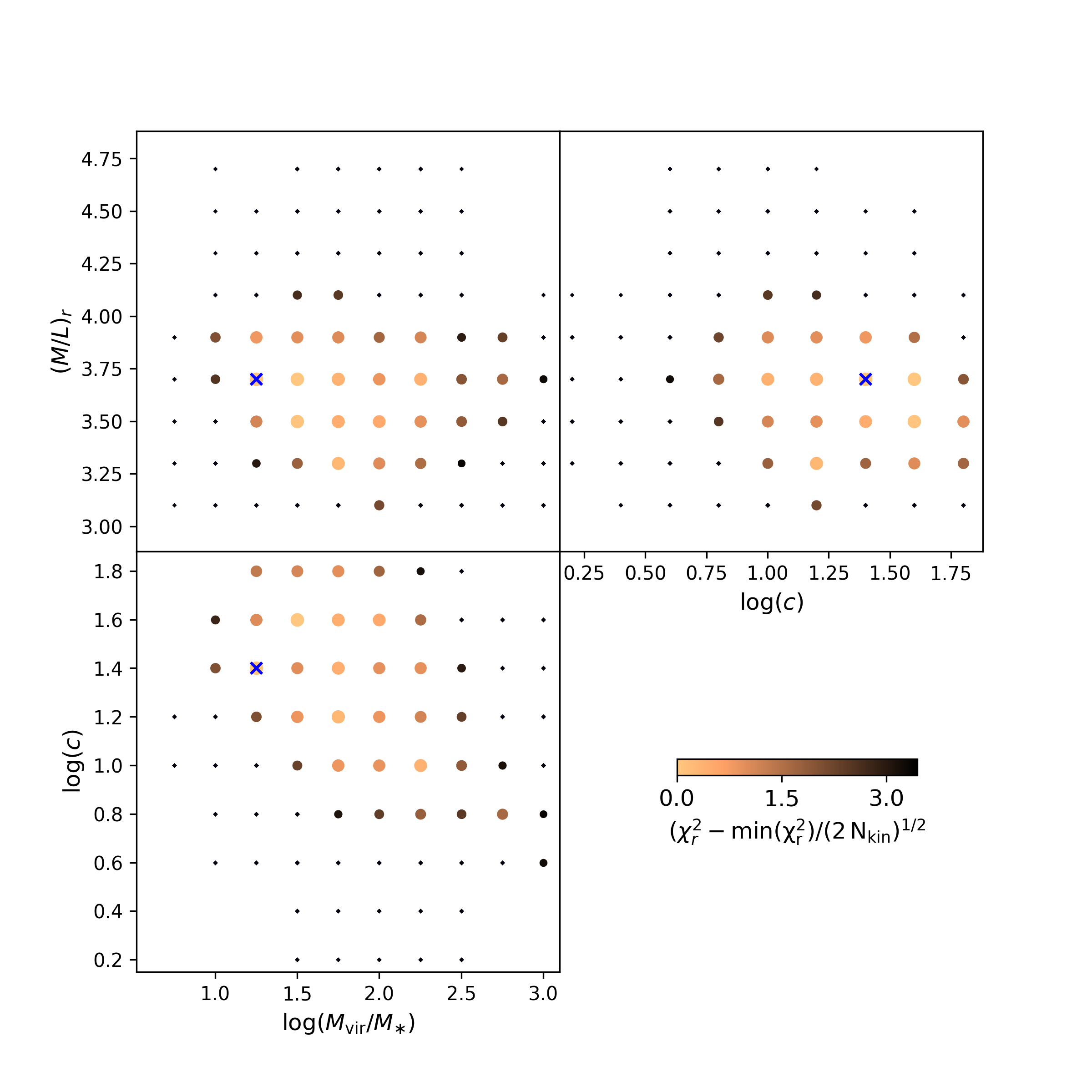}
    \caption{Illustration of the parameter grid explored by our model. Each dot represents one model. The color code and the symbol size give the reduced $\chi^2_r$, normalised such that the best fitting model as given by the blue cross reaches zero.}
    \label{fig:param_grid}
\end{figure}

\begin{figure*}
    \centering
    \includegraphics[width=0.99\textwidth]{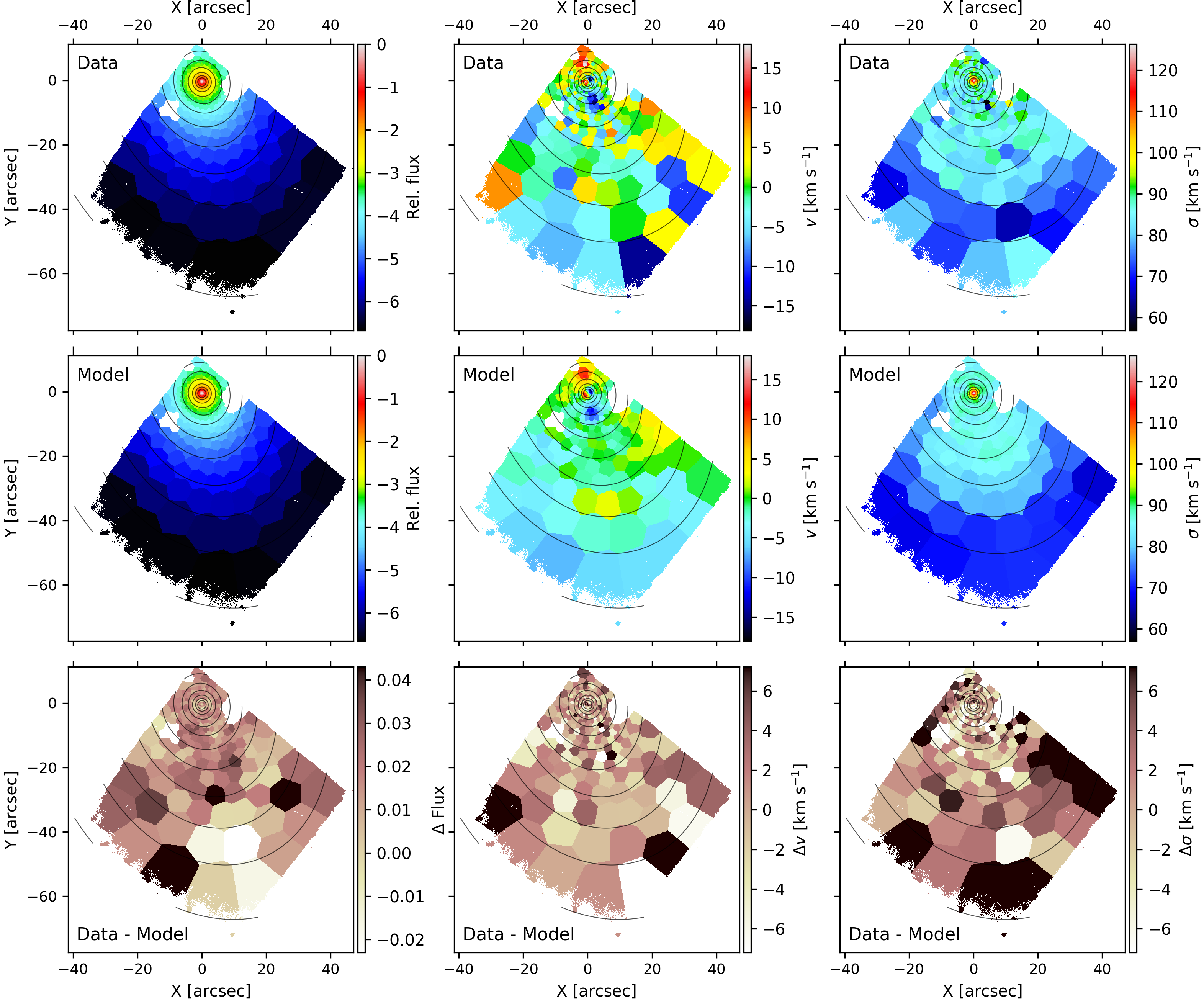}
    \caption{Best-fitting Schwarzschild model of FCC\,47. Top row: Voronoi binned kinematic data for FCC\,47. Left to right: Surface brightness, radial velocity and velocity dispersion. Middle row: Best-fit Schwarzschild model. Bottom row: Residuals created by subtracting the model from the data. The white patches indicate masked bins. The surface brightness counters are taken from the MGE model (see also Fig. \ref{fig:MGE}). The plotting ranges are not identical to those shown in Fig. \ref{fig:FCC47_kin}.}
    \label{fig:schwarzschild_model}
\end{figure*}

\begin{figure}
\centering
\includegraphics[width=0.48\textwidth]{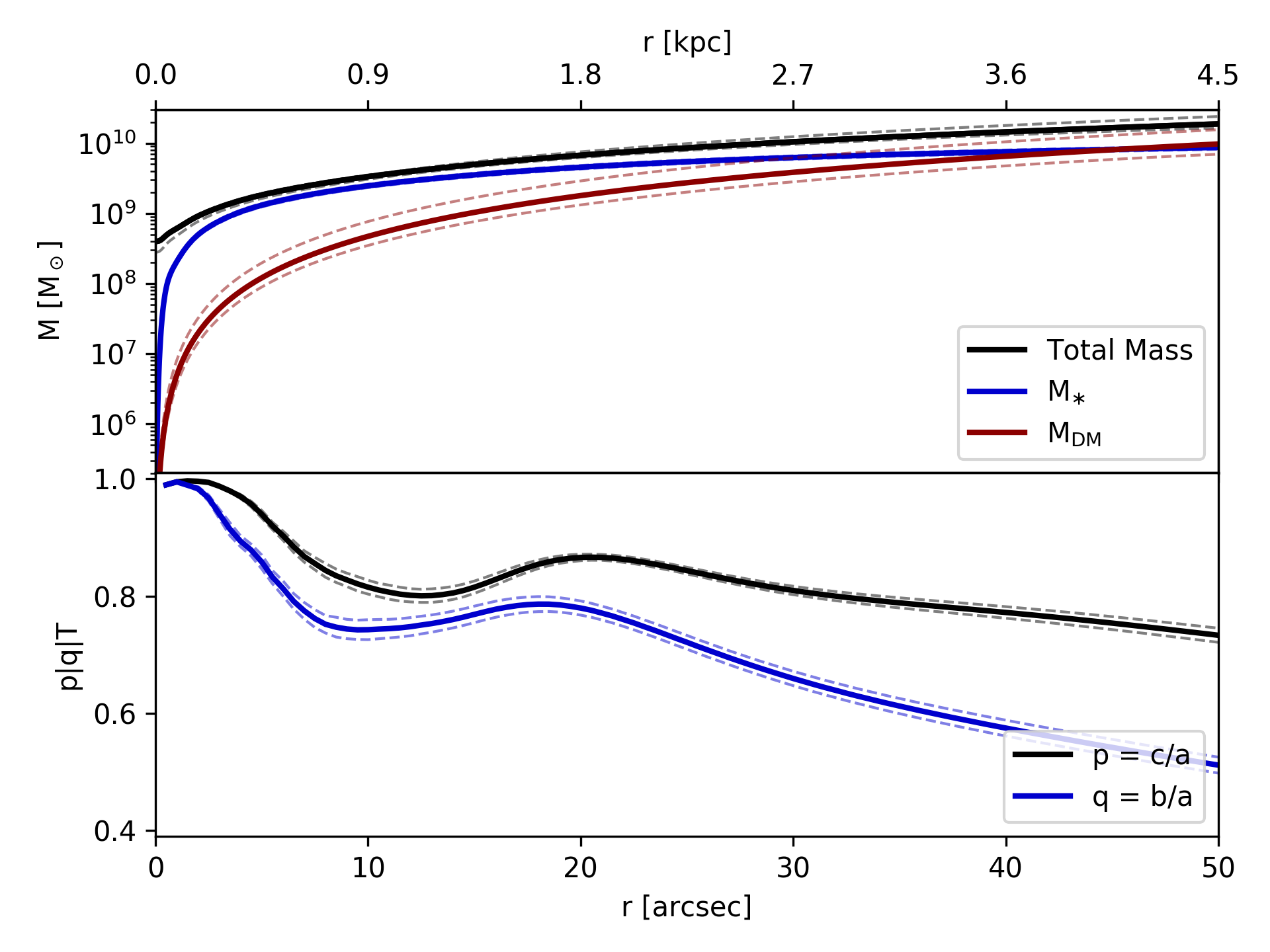}
\caption{Top: Profile of the enclosed mass for the stellar mass (blue), the dark mass (red) and the total (black). Bottom: Radial profile of the geometric shape parameters $q$ and $p$. The dotted lines indicate the 1$\sigma$ uncertainty determined from the five best-fitting models.}
\label{fig:schwz_profiles}
\end{figure}

\begin{figure}
\centering
\includegraphics[width=0.48\textwidth]{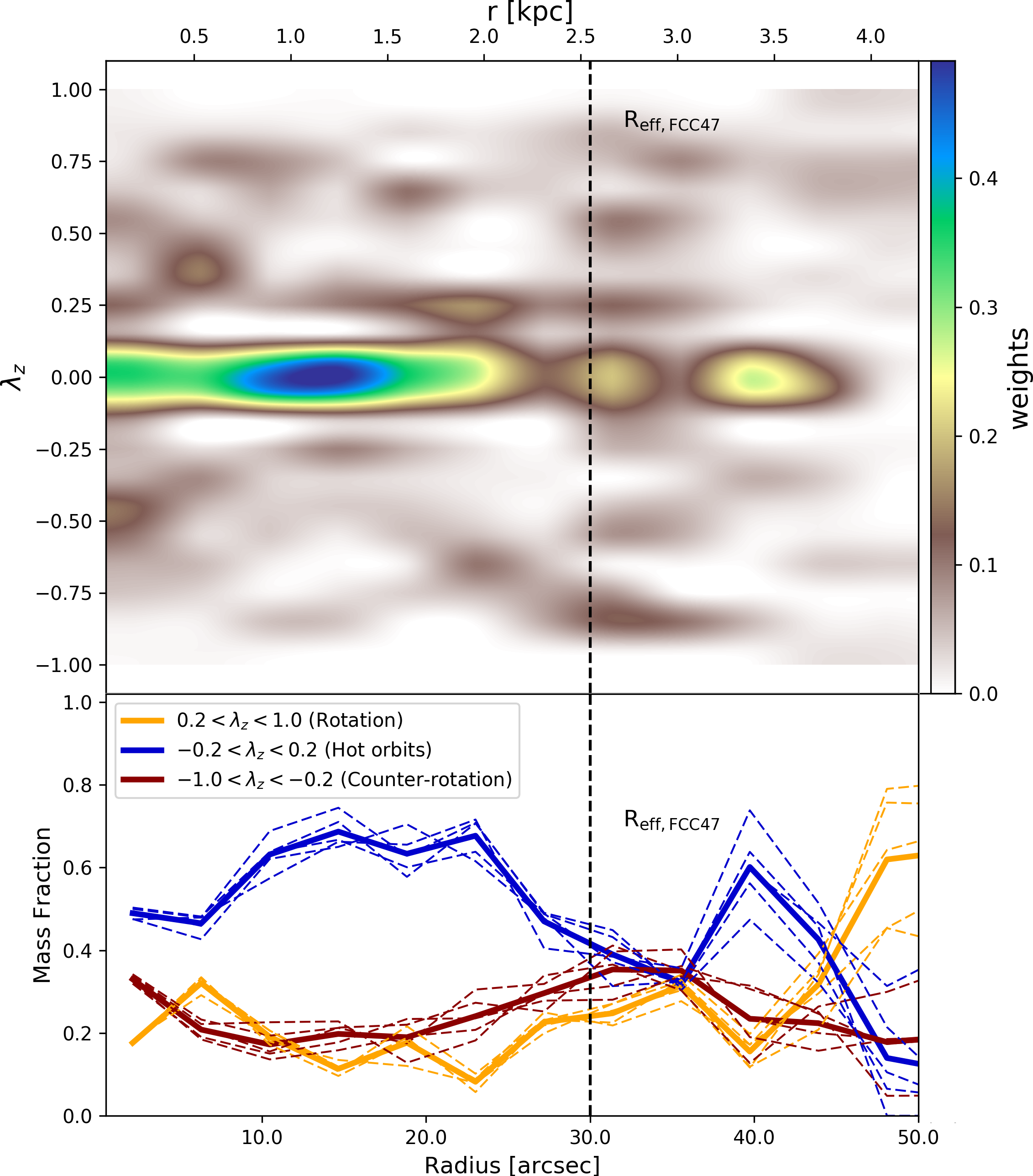}
\caption{Top: Distribution of orbital weights along short axis (top). We show a combination of the five best-fitting models. The dotted lines show the effective radius FCC\,47 ($R_\text{eff, FCC\,47} = 30\arcsec$, \citealt{Ferguson1989}). Bottom: Profile of the mass fractions of rotating (yellow), hot (blue) and counter-rotating orbits (red). The solid lines show the combination of the five best-fitting models (dashed lines).}
\label{fig:bestfit_orbit}
\end{figure}

\begin{figure}
\centering
\includegraphics[width=0.48\textwidth]{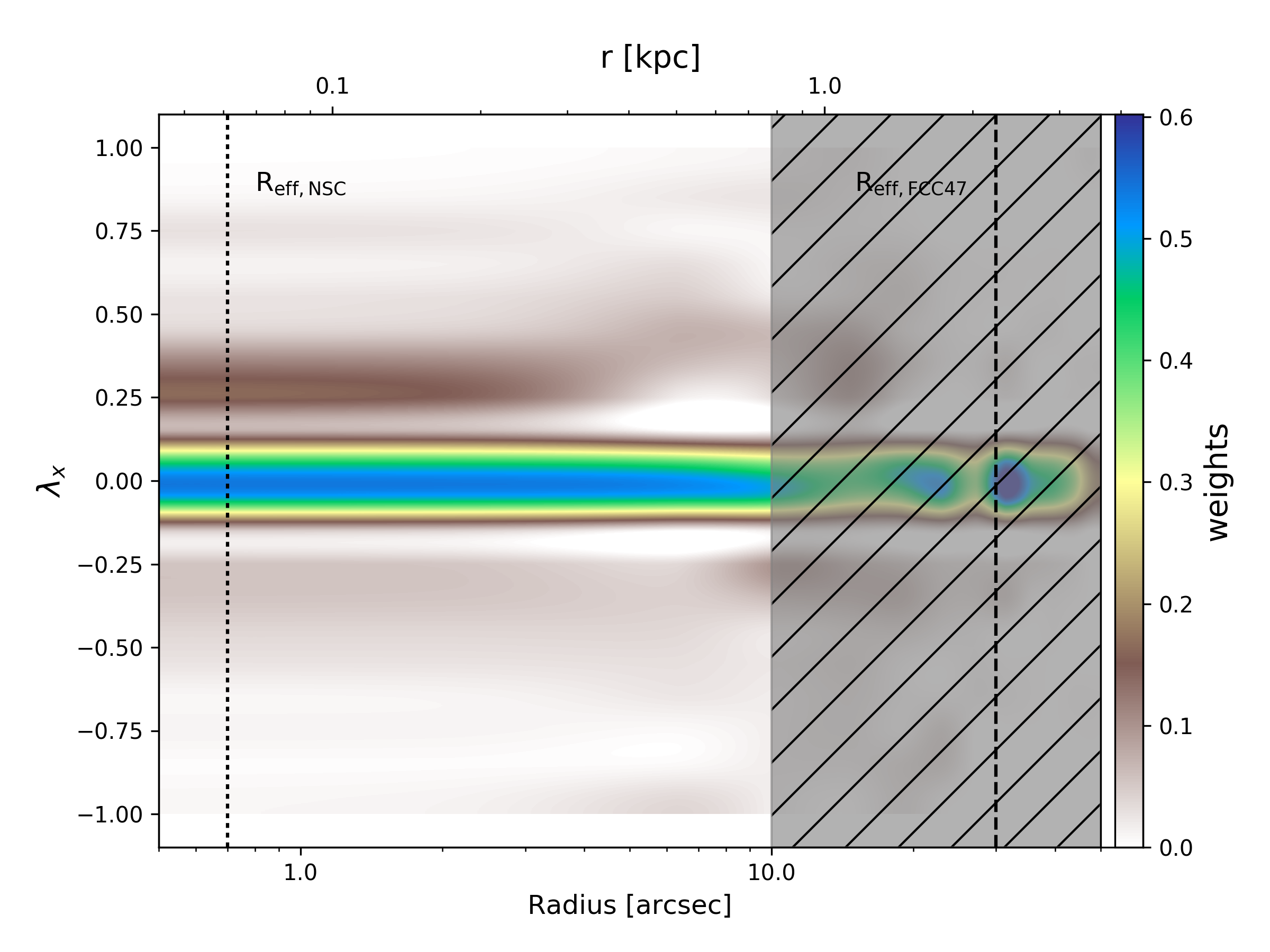}
\caption{Distribution of orbital weights along long axis of the mean of the five best-fitting models. The dotted lines show the effective radius of the NSC ($R_\text{eff, NSC} = 0.7\arcsec$, \citealt{Turner2012}) and FCC\,47 ($R_\text{eff, FCC\,47} = 30\arcsec$, \citealt{Ferguson1989}). The shaded area shows the region where we lack coverage with the MUSE data along the short axis.}
\label{fig:NSC_orbits}
\end{figure}

\subsection{Input to the model}
The gravitational potential of FCC\,47 was modelled by a combination of luminous (stellar) and dark matter (black hole + halo) distributions. The dark matter halo is described by the parameters listed above while the intrinsic stellar mass distribution was inferred from the observed stellar light using a MGE parametrisation of the 2D luminosity distribution. The photometric data should be acquired in the same wavelength regime from which the kinematics were extracted. This ensures that both photometry and kinematics describe the properties of the same tracer population. We used VST $r$-band data of FCC\,47 from the Fornax Deep Survey\footnote{FDS is a large scale survey of the Fornax galaxy cluster with the VLT Survey Telescope (VST) that covers $\sim$ 26 square degrees around the central galaxy NGC\,1399 in $u$, $g$, $r$ and $i$-bands} (FDS, \citealt{Iodice2016}) for the MGE parametrisation. The VST image has a pixel scale of 0.261\arcsec\,pix$^{-1}$ and a PSF FWHM of 0.87\arcsec. These values are comparable to the MUSE data (we assume a PSF FWHM of 0.7\arcsec), but the VST image covers the full galaxy instead of only one quarter, which was helpful to generate a better MGE parametrisation. Fig. \ref{fig:MGE} represents our MGE model.

For the kinematic maps, realistic uncertainties are required for each bin. To acquire these, we re-extract the LOSVD from the binned cube using the MILES models and 100 MC samples in each bin. The so created kinematic maps are consistent with those shown in Fig. \ref{fig:FCC47_kin} within the uncertainties that can reach $>$ 15 km s$^{-1}$ in the outermost bins.
As Schwarzschild models are inherently point-symmetric around the galaxy centre, we mask the signature of the UCD and the foreground star near the galaxy centre from the kinematic maps. We did not alter the bins where GCs are located as their contribution to their bins seems to be negligible.

\subsection{Schwarzschild model of FCC\,47}
We show the parameter grid of $M/L$ in the $r$ band, concentration log($c$) and dark matter fraction $\text{log}(M_{\text{vir}}/M_\ast)$ in Fig. \ref{fig:param_grid} to illustrate the explored parameter space. The color indicates the normalized $\chi^2$ and the best fitting model is marked by the blue cross. This model reproduced the rotational structures of the NSC and the disk component as shown in Fig. \ref{fig:schwarzschild_model}. We list the results of the model in Tab \ref{tab:schwz}. The viewing angles are well constrained by the kinematic features as only values within $\sim$ 10$^\circ$ around the best-fit values are allowed (see Fig. \ref{fig:angles} in the Appendix). The model for the radial velocity map also shows three bins at $\sim$ 30\arcsec distance from the centre with slightly higher velocities ($\sim$ 5 km s$^{-1}$) than the surrounding bins. The same feature is present in the observed data; however, comparing to photometry we cannot find any indication that this a real feature and the low velocity is fully consistent with the systemic velocity of FCC\,47, given that the uncertainties in these outer bins can be $>$ 15 km s$^{-1}$.

\begin{table}
    \centering
    \begin{tabular}{l l l} \hline
    Property & Value & Description \\ \hline \hline
    $\theta$ ($^\circ$) & 42 $\pm$ 2 & Viewing angle (Inclination) \\
    $\phi$ ($^\circ$) & 70 $\pm$ 1 & Viewing angle \\
    $\psi$ ($^\circ$) & 102 $\pm$ 2  & Viewing angle \\
    $M/L_r$ ($M_\sun/L_\sun$) & 3.5 $\pm$ 0.1 & $r$-band mass-to-light ratio \\
    $M_\ast$ ($10^{10} M_\sun$) & 1.5 $\pm$ 0.1 & Total stellar mass \\
    $M_\text{vir}$ ($10^{12} M_\sun$) &  1.1 $\pm$ 1.0 & Virial mass \\\hline
   \end{tabular}
       \caption{Results from the dynamical Schwarzschild model.}
    \label{tab:schwz}
   \end{table}

The enclosed mass profile including stellar and dark mass is shown in the top panel of Fig. \ref{fig:schwz_profiles}. Our model finds a total stellar mass of FCC\,47 of $M_\ast =1.5 \pm 0.1 \times 10^{10} M_\sun$, extrapolated outside of our data coverage. The derived value value is higher than what was found by \cite{Saulder2016}, who used photometric integrated magnitudes calibrated with a large sample of spectroscopic SDSS masses, a Kroupa IMF and $K$-band mass-to-light ratios to estimate the stellar mass to be $M_\ast = 6.3 \times 10^9 M_\sun$. They did not give a value for the uncertainty but note that the estimate of the photometric stellar mass should be possible within a factor of two. \cite{Liu2019} quoted a stellar mass of $M_\ast = 9.3 \times 10^9 M_\sun$, based on ACSFCS measurements originally used in \cite{Turner2012}. These measurements were made using ($g - z$) colours and the relations from \cite{Bell2003}. We find a dark matter fraction of $\text{log}(M_{\text{vir}}/M_\ast)$ = 1.5 and a total mass of the dark halo $M_\text{vir} = 1.1 \pm 1.0 \times 10^{12} M_\sun$. The uncertainty is large due the lack of data at larger radii. The bottom panel of Fig. \ref{fig:schwz_profiles} shows a radial profile of the shape parameters $p$ and $q$.

\subsection{Orbit distribution}
The distribution of orbital weights of our model is presented in Figure \ref{fig:bestfit_orbit} and Fig \ref{fig:NSC_orbits}. The distribution is shown on the phase space of circularity $\lambda_i$ and the intrinsic 3D radius $r$ of a given orbit (see \citealt{Zhu2018a}). We combined the orbit distributions of the five best fitting models.
The circularity is used to differentiate between different orbits: 
$\left|\lambda_i\right|$ = 1 indicates a circular (cold) orbit, while $\lambda_i = 0$ indicates a radial (hot) orbit. The sign refers to the sense of rotation, and thus negative values indicate counter-rotating orbits. 
Figure \ref{fig:bestfit_orbit} shows the distribution for orbits projected along the short axis $z$ as well as around the long axis $x$. While $z$ is aligned with the kinematic axis of the rotating disk and main photometric angle of the galaxy, the long axis $x$ is perpendicular. Although due to projection effects, the apparent angle between NSC rotation and the disk is $\sim$ 115$^{\circ}$, stable orbits only exist around the long or short axis.

The orbital distribution around the short axis (Fig. \ref{fig:bestfit_orbit}) is complex. In the inner region, the orbital structure is dominated by hot and warm orbits ($\mid\lambda_z\mid < 0.7$). Over all radii, the fraction of counter-rotating orbits is high, as also the radial plot of mass fractions of the five best fitting models in the bottom panel of Fig. \ref{fig:bestfit_orbit} shows. This explains the negligible rotation in the outskirts of the galaxy as the two counter-rotating populations with warm orbits cancel each others rotation signatures while still showing a high velocity dispersion. This kinematic feature has been routinely observed in galaxies hosting extended counter-rotating stellar components (see \citealt{Corsini2014} for a review).

The distribution for the circularity around the long axis shows less structure. It is completely dominated by hot box orbits over all radii. The warm orbits at small radii ($\lambda_x \sim 0.25, r \lesssim 3\arcsec$) highlight the rotation of the NSC. The radial extent of these weights are larger than the effective radius of the NSC, but match the apparent size of the NSC in the MUSE data ($\sim$ 3\arcsec). The spots with small weights at r $>$ 10\arcsec\,are most likely not connected to real features. Instead, they are an effect of the limited size of the orbit library and of the limited coverage of the MUSE data along the short axis that only extends out to $\sim$\,10\arcsec\,along this axis.

The orbital distributions shown in Fig. \ref{fig:bestfit_orbit} reveal the nature of the two kinematically decoupled components in FCC\,47. The rotation of the disk seen on a $\sim$ 2 kpc scale (see Fig. \ref{fig:FCC47_kin}) is not a decoupled component as such, instead the high fraction of counter-rotating orbits indicate that this apparent rotating disk is the result of two much more extended counter-rotating populations that show an imbalance on the scale of the disk. Interestingly, according to our model, we are seeing the counter-rotating population that has a larger mass fraction by a few percent on the extend of the disk. The small rotation amplitude of the apparent disk rotation can be explained by this small difference in mass fractions. On the other hand, the NSC appears to be a separate component with a distinct kinematic signature. Besides the NSC, there is no evidence for other populations that show disk-like prolate rotation. 
The limited spatial resolution of the MUSE data does not allow to estimate a contribution from a central SMBH to this mass. 

\section{Discussion}
\label{sect:discussion}
FCC\,47 is a  ETG with $M_\ast \sim 10^{10} M_\sun$ in the outskirts of the Fornax cluster with a particularly large NSC and rich GC system. Our MUSE study has revealed several peculiarities that distinguish this galaxy from others, for example the presence of two visible KDCs. In the following we use our comprehensive collection of information to put constraints on the formation history of FCC\,47's massive NSC.

\subsection{The star clusters in FCC 47}
\subsubsection{The nuclear star cluster}
FCC\,47 was targeted because of its particularly large NSC with an effective radius of 66 pc \citep{Turner2012}. 
Using a large sample of NSCs in early- and late-type galaxies, \cite{Georgiev2016} studied various scaling relations between mass and size of the NSC and host mass properties. The observed relation between NSC size and host mass would predict a NSC with an effective radius of $\sim$\,20\,pc for a galaxy with the stellar mass of FCC\,47. However, the observed radius of 66 pc is still within the scatter and there are several other ETGs with NSCs of similar sizes in this mass range. In addition to the size, our data allowed to estimate the stellar mass of the NSC to be $\sim 7 \times 10^8 M_\sun$ using stellar population models. This is a massive NSC and is slightly more massive than NSCs of a similar size, but again it lies within the scatter of the NSC size-mass relation \citep{Georgiev2016}. Nonetheless, this high mass places the NSC among the most massive known NSCs (e.g. \citealt{SanchezJanssen2018}).

The NSC constitutes the peak of metallicity and stellar age in FCC\,47 and is clearly more metal-rich than the immediate surrounding galaxy, in fact it is possible that a large fraction of the strong central metallicity gradient can be explained by the metal-rich NSC being smeared out by the MUSE PSF. While many NSCs were found to also contain young stellar populations \citep{Rossa2006, Paudel2011}, the NSC in FCC\,47 shows no contribution from young stars ($<$ 10 Gyr). It contains the oldest stellar ages, indicating a quick metal-enrichment and early quenching of star formation. We could identify two subpopulations in the NSC, a dominant population with super-solar metallicities and intermediate [$\alpha$/Fe] and a secondary component that has a lower metallicity ($\sim -0.4$ dex) and a higher alpha-abundance ratio. These two populations could be explained by continuous and efficient self-enrichment of an initially metal-poor and alpha-enhanced stellar population that increased metallicity and decreased [$\alpha$/Fe] due to the pollution from supernovae Ia ejecta. These supernovae release iron into the interstellar medium within $\sim$ 1 Gyr \citep{Thomas2005}, in agreement with the exclusively old ages we found in the NSC. As we could not find different SFHs for the metal-poor and metal-rich subpopulations, the self-enrichment must have stopped early in this picture. Alternatively, the more metal-poor component could indicate NSC mass growth through the accretion of GCs of this chemical composition. 

In the study of \cite{Lyubenova2019} that used high angular resolution IFU SINFONI data of five NSCs of galaxies in the Fornax cluster, the NSC of FCC\,47 stood out with having the strongest rotation and consequently having the highest angular momentum of the observed sample. 
The MUSE data confirmed the rotation of the NSC, but allowed us to set it into context with the rest of the galaxy and we see that the NSC rotates as a KDC around an axis offset by 115$^\circ$ from the main photometric (and rotation) axis of the galaxy. While the NSC in the Milky Way shows evidence for a small misalignment between the rotation of the NSC and the galactic plane \citep{Feldmeier2014}, the gas-rich late-type galaxies studied by \cite{Seth2008b} have NSCs that rotate in the plane of their host. 
Due to lower intrinsic rotation and a more spherical geometry, kinematically misaligned NSCs could be easier to spot in ETGs.

The NSC in FCC\,47 was originally identified as an excess of light at the centre of the galaxy \citep{Turner2012}. Our analysis showed that the NSC is indeed a distinct component of FCC\,47, both in its kinematic and stellar population properties. 

\subsubsection{The globular clusters}
FCC\,47 has a rich GC system with more than 300 candidates identified in the ACSFCS \citep{Jordan2015}. We could identify 42 GCs in our MUSE data, with 25 having a spectral S/N high enough to extract their LOS velocity. With this small number GC sample, we could not identify rotating substructures, but we found that the GCs show a large velocity dispersion. This is in agreement with other studies that have established that GCs usually are a dynamical hot system (e.g. \citealt{KisslerPatig1998, Dabringhausen2008}), supported by velocity dispersion. However, there have been detections of rotating subsystems in other galaxies, e.g., within the SLUGGS survey \citep{Foster2016, Forbes2017}. We noted that the blue GC subpopulation shows a larger velocity dispersion than the red GCs, as in most other galaxies (e.g. \citealt{Schuberth2010}). In addition, the ACSFCS study shows that red GCs are more centrally concentrated and as a consequence the distribution of GC colours shows a radial gradient from red to blue in the outskirts. 

From the five GCs that are bright enough to estimate their metallicity, we found four blue GCs that are significantly more metal-poor than the galaxy at their projected distance. The fifth GC is red and has a higher metallicity, similar to the NSC. Combined with the fact that we could not find indications of young ages in the GCs or the stellar light, the bimodality of GC colours in FCC\,47 might be truly an effect of a metallicity bimodality, however, we only have five GCs to support this. Using the photometric colours from the ACSFCS and assuming an old age (13 Gyr) for all ACSFCS GC candidates, we estimated that the GC system has a total mass of $\sim 1 \times 10^8 M_\sun$. This is only $\sim$ 17\% the mass found in the NSC. 

The larger velocity dispersion, the low metallicities of the blue GCs and the GC colour gradient indicate an ex-situ origin of the blue GC system. Generally, the bimodal colour distribution of GCs as observed in many galaxies is interpreted in this scheme, with the red (metal-rich) GC population having formed in-situ while the blue GCs were accreted through minor mergers with metal-poorer dwarf galaxies \citep{Cote1998, Hilker1999, Lotz2004, Peng2006, Cote2006, Georgiev2008}. The red, metal-rich GC population might have formed together with the NSC and the metal-rich galaxy population, that has a smaller mass fraction and thus is hidden within the metal-poor population of FCC\,47, that is particularly dominant at larger radii.

\subsection{Constraints on the formation of FCC 47's nuclear star cluster}
Having collected information on the kinematic and stellar population structure of FCC\,47 itself, its NSC and the GCs, we want to explore if we can put constraints on the formation of FCC\,47's massive NSC by exploring the two most commonly discussed scenarios and their implications.

\subsubsection{Globular cluster accretion scenario}
As discussed in the introduction, in the GC accretion scenario, the NSC forms from the (dry) accretion of in-spiralling GCs that make their way to the centre due to dynamical friction. 
Although the in-fall of GCs with random orbits can result in a low net angular momentum of the NSC, simulations have shown that some NSCs can have significant rotation. In the case of the NSC of FCC\,47, a comparison to $N$-body simulations \citep{Antonini2012, Perets2014, Tsatsi2017} has shown that this high angular momentum can still be explained by gas-free accretion of GCs, but requires a preferred in-fall orbital direction as discussed by \cite{Lyubenova2019}. However, we could not find any indication for rotation substructure in the GC population, possible due to the low numbers of available GC LOS velocities and an incomplete spatial coverage. In a future wide-field study of the kinematics of the red GC subpopulation that might share the high metallicity of the NSC could give a handle on the contribution of GC accretion to the NSC mass build up. However, the subsequent evolution of the cluster system and merger events of the host galaxy could have removed any rotation substructure within the GCs. 

The old ages and the absence of young populations in the NSC are consistent with the accretion of gas-free GCs. We find that the NSC constitutes the metal-rich peak of FCC\,47 and is significantly more metal-rich than the blue (metal-poor) GC population, however, we find at least one red GC that has a similar metallicity. Our current sample of GCs with metallicity measurements is very limited, but the comparison of GC colours shows that there probably still are GCs in FCC\,47 with metallicities similar to the NSC. These red GCs are also more centrally concentrated and thus it is possible that the NSC has formed or grown by accreting such GCs, similar to the GCs found in the bulge of the MW (e.g. \citealt{Cote1999, Munoz2017, Munoz2018}). The presence of a minor more metal-poor population in the NSC, might indicate some accretion of old, metal-poor and alpha-enhanced stars, possibly also accreted in an in-spiral of a few metal-poor GCs. However, it is unclear if the steep central metallicity gradient can be explained by GC in-spiral because the disk region already shows significantly lower metallicites or whether (and how much) additional gas accretion would be needed. 

Similarly, the composite formation scenario proposed by \cite{Guillard2016} could give a viable explanation for the formation of the NSC in FCC\,47. In their scenario, a massive gas-rich GC forms close to the centre and spirals inward within a few Gyr, possibly followed by a merger with a second gas-rich GC. Depending on whether a second GC falls to the centre, this model predicts significant rotation in the formed NSC and a fast quenching of star formation in the centre because of gas being expelled, in agreement with the old age and high metallicity we find for the NSC in FCC\,47. However, it is unclear if this scenario can explain the KDC with the offset rotation axis and the described accretion of only two massive GCs is not viable for FCC\,47. To explain its mass, a large number of star clusters must have merged within the first Gyr of the galaxy.

Independent of the kinematic and chemical structure of the NSC and the GCs, the high mass of the NSC makes formation solely by GC-accretion unlikely. As our estimate shows, the NSC is much more massive than the total GC system as observed today. To explain its mass of $\sim$ 7 $\times 10^8 M_\sun$, it must have accreted early-on hundreds of GCs that must have been quite massive ($> 10^6 M_\sun$) to explain the high (and homogeneous) metallicity. Modelling the effects of dynamical friction, evolution including mass loss and tidal stripping on the GC system of the MW and M\,87, \cite{Gnedin2014} found that GC accretion can contribute a large fraction of mass to NSCs in galaxies with $M_\ast < 10^{11} M_\sun$. However, following their predicted scaling between NSC and galaxy stellar mass, FCC\,47's NSC should have a mass of \mbox{$\sim 1 \times 10^8 M_\sun$} assuming growth through GC accretion. This estimate would increase if FCC\,47 had been extremely efficient in forming massive metal-rich star clusters early on, but it is questionable how such a large population of metal-rich GCs could have formed. 

We therefore conclude that the formation of the NSC in FCC47 is unlikely to have been dominated by the gas-free accretion of old GCs. However, from simple dynamical arguments, it follows that there must have been some GC accretion in such a rich GC system and the accretion of young metal-rich star clusters early-on is still possible. Detailed modelling that accounts for the properties of FCC\,47's GC system would be needed to get a handle on the relative fraction of mass build up via dry GC accretion.  

\subsubsection{In-situ formation}
The formation of the NSC through in-situ formation from infalling gas can explain the high mass and high metallicity better, but requires that a large supply of gas has been funnelled originally into the centre of FCC\,47. This supply must have been quickly exhausted after a short period of efficient star formation and self-enrichment. With our data, we cannot constrain the mechanisms that have stopped the star formation in the NSC and the rest of the galaxy. We do not know if the gas was simply exhausted or it was removed, possibly due to interactions in the Fornax cluster environment, stellar feedback or feedback from an active galactic nucleus.

As the study of \cite{Lyubenova2019} has shown, the high angular momentum of the NSC is inconsistent with formation via the in-spiral of GCs from random directions, so would need coherent anisotropic accretion of GCs. Therefore, the in-situ formation from infalling gas might be a more viable explanation for the formation of the NSC in FCC\,47. In this scenario, the infalling gas would inherit the angular momentum from its origin. However, there is no indication of other stellar populations rotating around the same axis as the NSC, neither in the galaxy main body nor within the GC population, but a larger sample of (red) GC velocities is needed to confirm this. Instead, the rotation axis of the NSC is offset with respect to the photometric axis and main rotation axis of the galaxy. The origin of the kinematically decoupled rotation of the NSC might not be connected to intrinsic formation scenarios as discussed, but could be evidence for a major merger that has altered FCC\,47 kinematic structure significantly. Such a merger and the corresponding merger of the NSCs of the progenitor galaxies could also explain the high mass of FCC\,47's NSC.

\subsubsection{Evidence for merger history of FCC 47}
KDCs are not a rare phenomenon, especially in massive ETGs ($M_\ast > 10^{10.8} M_\sun$, \citealt{Krajnovic2013}). Classical KDCs have kpc-sizes, similar to the rotating disk structure in FCC\,47. Our dynamical model shows that the disk KDC is not truly decoupled, similar to the KDC in NGC\,4365 \citep{vanDenBosch2008, Nedelchev2019}. Conversely, the NSC is a distinct kinematic component and is in its size comparable to the class of compact KDCs with sizes of $\sim 100$ pc \citep{McDermid2007}. In contrast to the NSC of FCC\,47, these more compact KDCs usually have noticeably younger ages \citep{McDermid2006}, although the apparent size of the NSC in the MUSE data ($\sim 3\arcsec \approx 250$ pc) places it in between these two classifications. 

Simulations of galaxy mergers have been able to explain the formation of KDCs, but they always require a major merger between massive galaxies. The central kinematic decoupling arises if, for example, gas dissipation, retrograde merger orbits or the merger of gas-rich disks are incorporated in the simulations \citep{Balcells1990, BarnesHernquist1992, Jesseit2007, Bois2011}. \citet{Tsatsi2015} showed that kpc-sized KDCs can also form in a prograde merger of two disk galaxies during which reactive forces create short-lived reversal of the orbital spin. 

\cite{Rantala2019} shows that KDCs may also arise from orbit reversals of the central SMBH caused by gravitational torques from expelled material in the dissipationless merger of two massive ETGs. For binary mergers of massive ETGs (each $M_\ast >  8 \times 10^{10} M_\sun$) with very massive SMBHs (each $> 10^{9} M_\sun$), they even find multiple KDCs, similar to what we found in FCC\,47, however, the simulated KDCs still share a single rotation axis aligned with the minor axis of the galaxy. In case the described scenario scales down to lower-mass galaxies such as FCC\,47, the simulations predict a very massive SMBH as well as a tangential velocity anisotropy in the centre. Both the presence of a SMBH and a central tangentially biased velocity structure could not be tested with the MUSE data due to the limited spatial resolution but could be measured by improving the current dynamical orbit-based model by including high angular resolution data of the central region, e.g., incorporating SINFONI data into the existing model in a future study.

Although current simulations might not be sufficient to explain the KDCs in FCC\,47, the presence of the two apparent KDCs and the two counter-rotating populations seems to be strong evidence of at least one major merger. In particular, the presence of two decoupled components in combination with the exclusively old ages is intriguing. The kinematics could suggest a scenario where the NSC with its decoupled rotation has fallen into the already formed galaxy, perhaps as the remnant nucleus of a destroyed galaxy. However, following the mass-metallicity relation (e.g. \citealt{Gallazzi2005, Kirby2013}), the high metallicity of the NSC would then imply that it must have formed in an originally massive galaxy, similar to FCC\,47. Alternatively, the NSC might have formed as a red nugget at high redshift and would have ended up as a metal-rich compact elliptical (cE) or massive relic \citep{Ferre-Mateu2017, Martin-Navarro2019} but instead got captured by FCC\,47. This scenario could explain the metallicity as many cEs are known to have a relatively high metallicity for their mass \citep{Zhang2018}.

In comparison to other galaxies of the Virgo and Fornax cluster, FCC\,47 not only has a large and massive NSC \citep{Turner2012}, but also has a higher specific frequency of GCs within one effective radius than other galaxies of similar mass \citep{Liu2019}. This might imply that the formation of the NSC and the red, centrally concentrated GC system is connected and thus a scenario in which the NSC is an original part of the galaxy might be favoured. In such a scenario, the NSC, the metal-rich part of galaxy and the red, metal-rich GCs could have formed together in clumpy, bursty star formation at high redshift (see also \citealt{Baesley2018}). The galaxy then later underwent at least one major merger that created the counter-rotating populations and younger, more metal-poor outskirts. Specialized simulations would be required to test whether the decoupled rotation of the NSC could have survived the subsequent evolution of the galaxy or whether the kinematic decoupling itself could be a result of second major merger.

Independent of the exact formation scenario that has caused the peculiar kinematic structure of FCC\,47, the presence of KDCs indicate the importance of mergers in the formation of both galaxy and NSC. Thus, studying other massive NSCs and their hosts is required to determine if the commonly discussed NSC formation scenarios are sufficient to explain the formation of the most massive NSCs in general.

\section{Conclusions}
\label{sect:conclusion}
In this paper, we presented a comprehensive study of the kinematic and stellar population structure of the early-type galaxy FCC47 using MUSE WFM AO SV data. We have analysed both the integrated stellar light and GCs as discrete tracers of the kinematic and chemical structure, connecting the results to the kinematic and chemical study of the NSC. Its particularly large size allowed us to resolve its structure. We summarize our results as follows:
\begin{itemize}
\item We found that FCC\,47 shows a complex velocity structure with two KDCs on different scales. While there is no significant rotation on large scales, a rotating disk structure with a rotation amplitude of $\sim$ 20 km s$^{-1}$ is found with an extent of $\sim$ 2 kpc. In addition, our MUSE observation confirms the high angular momentum, strong rotation and high velocity dispersion of the NSC as discovered in earlier work. The rotation axis of the NSC and of the inner disk are offset by $\sim$ 115$^\circ$, constituting a second KDC. 

\item We constructed a Schwarzschild model to reveal FCC\,47's orbital structure.
    According to this model, the main body of FCC\,47 consists of two counter-rotating populations. On large scales, the counter-rotation cancels out any rotation signature. Near the centre, the two counter-rotating populations show an imbalance, thus creating the emergence of a inner rotating disk structure. The NSC appears to be truely kinematically decoupled and we find no indication of other populations showing the same sense of rotation. The Schwarzschild model determined the total stellar mass of FCC\,47 to be M$_\ast = 1.6 \times 10^{10} M_\sun$.
    
 \item FCC\,47 is overall old ($>$ 8 Gyr) and shows an age gradient from very old ages in the centre to younger ages in the outskirts. We did not find any evidence of young populations, gas or dust. While the galaxy is metal-poorer on larger scales, we find that the centre reaches solar metallicities and shows a strong central gradient that flattens outwards. The region of high metallicity matches the extent of the NSC rotation structure. We therefore assumed that the high metallicity is associated to the NSC while the rest of the galaxy is significantly more metal-poor with a shallow metallicity gradient. The NSC does not stand out in the map of light element abundance ratios ([$\alpha$/Fe]).
    
   \item Dividing FCC\,47 into the NSC, the (rotating) disk and the main galaxy body showed that the NSC is dominated ($\sim$ 70 \% in mass) by a metal-rich, old stellar population. This mass fraction decreases down to 47 \% in the disk and to $<$ 30 \% in the outskirts, where the galaxy is dominated by a more metal-poor population. 

\item Studying the galaxy-subtracted NSC spectrum confirmed the old age of the NSC, but we found two chemically distinct subpopulations: a dominating (67 \%) metal-rich population with intermediate light element abundance ratio and a secondary population with lower metallicity and higher [$\alpha$/Fe]. From the stellar population analysis, we found that the NSC is very massive ($M \sim 7 \times 10^8 M_\sun$).

\item We extracted MUSE spectra of 42 GCs using an ancillary catalogue from \cite{Jordan2015} as reference. 25 of these have a sufficient spectral S/N to determine their LOS velocity and a subsample of five GCs is bright enough for an estimate of their metallicities.

\item Within our sample of GC velocities, we did not see any rotation, but the blue GC population shows a larger velocity dispersion than the red population. Four of the five GCs that allowed a measurement of their metallicity have blue colours and are found to be significantly metal-poorer than the galaxy, while the fifth GC is associated with the red GC population and has a higher metallicity similar to the NSC. The red GCs could have formed together with the NSC and the metal-rich stellar population. Based on photometric metallicities, we estimated the total mass in GCs to be $\sim 1 \times 10^8 M_\sun$. 

\item We interpreted our results with respect to the most discussed formation channels of NSCs: the gas-free GC accretion and the in-situ formation scenario. With its high angular momentum, high mass and high metallicity, the NSC of FCC47 is unlikely to have formed dominantly by accreting non-rotating, metal-poor GCs as we find them in FCC\,47 today. However, we could not rule out that the NSC has formed a fraction of its mass by accreting more metal-rich GCs or gas-rich massive star clusters early on. Possibly, the decoupled kinematics of the NSC are not an effect of the formation pathway, but rather are the result of a major merger that has altered FCC\,47's kinematic structure.

\end{itemize}
To constrain the relative contribution of the GC-accretion and in-situ formation channel, a comparison with specialised models is required. To accurately constrain the GC-accretion channel, a more in-depth view on the kinematics with more LOS velocities and a larger spatial coverage of the red GC population is required because these seem to share the high metallicity of the NSC. 

FCC\,47 was targeted as MUSE WFM + AO SV target because of its large, strongly-rotating NSC and the rich GC system. Being a ETG in the outskirts of the Fornax cluster, the wealth of complexity in this galaxy has been surprising: we found two KDCs, a sharp central metallicity peak and FCC\,47-UCD1. As we argue in \cite{Fahrion2019}, the UCD could be the stripped nucleus of a disrupted dwarf galaxy and would therefore indicate a minor merger in FCC\,47's past. In addition, the KDCs most likely are evidence of at least one major merger event that has altered the kinematic structure of the galaxy significantly.

\begin{acknowledgements}
We thank the anonymous referee for helpful comments and suggestions that improved this manuscript. This work is based on observations collected at the European Organization for Astronomical Research in the Southern Hemisphere under ESO programme 60.A-9192.
GvdV acknowledges funding from the European Research Council (ERC) under the European Union's Horizon 2020 research and innovation programme under grant agreement No 724857 (Consolidator Grant ArcheoDyn). J. F-B acknowledges support from grant AYA2016-77237-C3-1-P from the Spanish Ministry of Economy and Competitiveness (MINECO).
E.M.C. acknowledges financial support from Padua University through grants DOR1715817/17, DOR1885254/18 and BIRD164402/16. RMcD is the recipient of an Australian Research Council Future Fellowship (project number FT150100333). LZ acknowledges support from Shanghai Astronomical Observatory, Chinese Academy of Sciences under grant NO.Y895201009
\end{acknowledgements}

\bibliographystyle{aa} 
\bibliography{References}
\appendix
\section{Appendix}
\subsection{S/N requirements}
\label{app:SNR}
\begin{figure}
\centering
\includegraphics[width=0.49\textwidth]{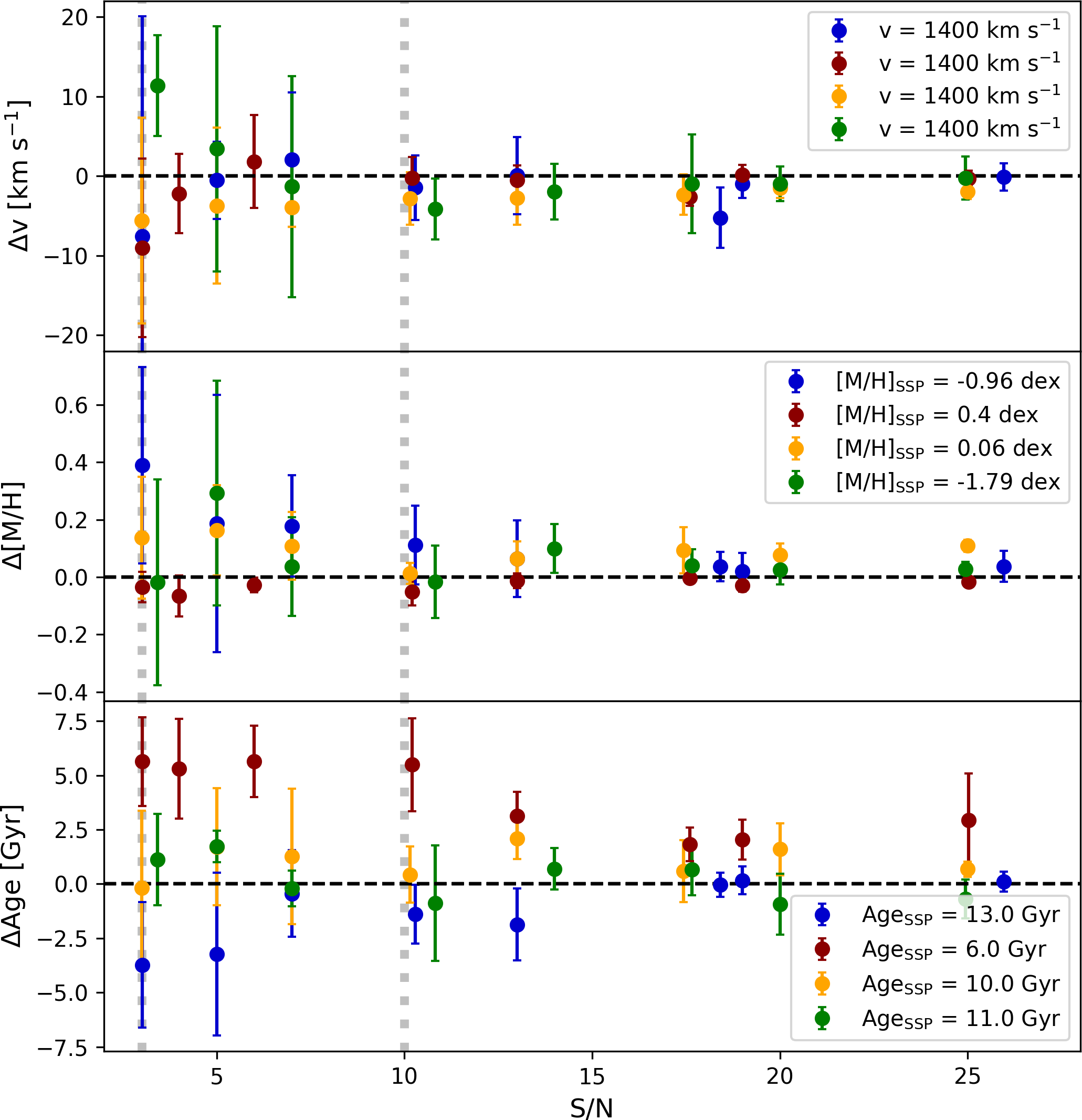}
\caption{Recovery of SSP input parameters with \textsc{pPXF} versus S/N of the spectrum. We use four different SSP templates (as indicated by different colours) from the E-MILES library and add artificially Gaussian noise. The noisy spectrum is then fitted with \textsc{pPXF}. For each S/N, we repeat this with five noisy spectra and show the difference between the input parameter and the mean values of these five trails and their standard deviation. The vertical lines indicate the threshold S/N values of 10 and 3 that we chose for reliable measurements of metallicities and velocities, respectively.}
\label{fig:SNR_testing}
\end{figure}

In order to verify that we can extract stellar kinematic and population properties from the low S/N spectra of the GCs, we set up several tests that use the E-MILES SSP templates as test spectra with known LOSVD parameters, metallicity and age. We cut the spectra to the MUSE wavelength range and artificially redshift and broaden it to a certain radial velocity and velocity dispersion. Then, random noise is added to reach a specific S/N before fitting the spectrum with \textsc{pPXF}. Figure \ref{fig:SNR_testing} illustrates how well velocity, age and metallicity are recovered by \textsc{pPXF} as a function of the spectral S/N. This is shown with four different SSP templates as input and at each S/N value, the fit is repeated five times to estimate the mean and standard deviation. 

We find that a S/N $\geqslant$ 5 is needed to get a radial velocity estimate within 10 km s$^{-1}$ accuracy. A S/N $\geqslant$ 3 is required to get radial velocities within $\sim$ 20 km s$^{-1}$ accuracy. This should be sufficient to determine the velocity dispersion of a large sample of GCs or identifying rotating substructure. Testing with different metallicities, we find that the uncertainties of the radial velocities increase for more metal-poor spectra. This is most likely caused by the lack of strong lines, which makes fitting for the velocity with \textsc{pPXF}'s cross-correlation method more difficult.

We find a S/N $\geqslant$ 10 is required to determine the metallicity within 0.2 dex. The age is less well recovered at these low S/N, in particular young ages. 

\subsection{Multi-Gaussian expansion model of FCC\,47}
\begin{figure}
\centering
\includegraphics[width=0.49\textwidth]{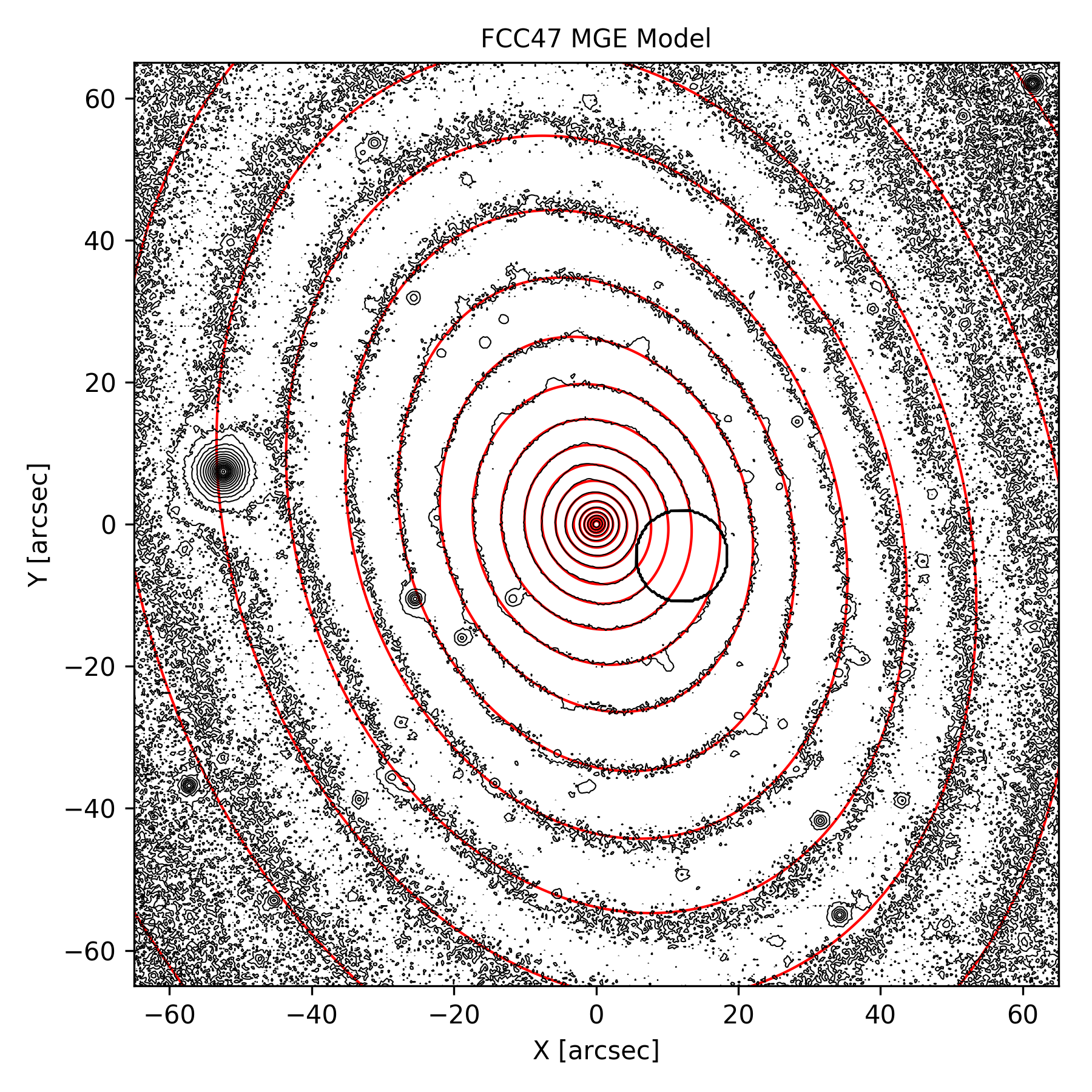}
\caption{Illustration of the MGE model of FCC\,47 used in our Schwarzschild model. The black contours show the original VST image, the red contours indicate the MGE model. We mask the foreground star close to the galactic centre.}
\label{fig:MGE}
\end{figure}

We illustrate the MGE model of FCC\,47 as used in our Schwarzschild model (Sect. \ref{sect:schwarzschild}) in Fig. \ref{fig:MGE}. The model is based on a $r$-band VST image of FCC\,47. The original data with some masked sources is shown in black while the contours of the MGE model are overplotted in red.

\subsection{Parameter grid of viewing angles}
\begin{figure}
\centering
\includegraphics[width=0.49\textwidth]{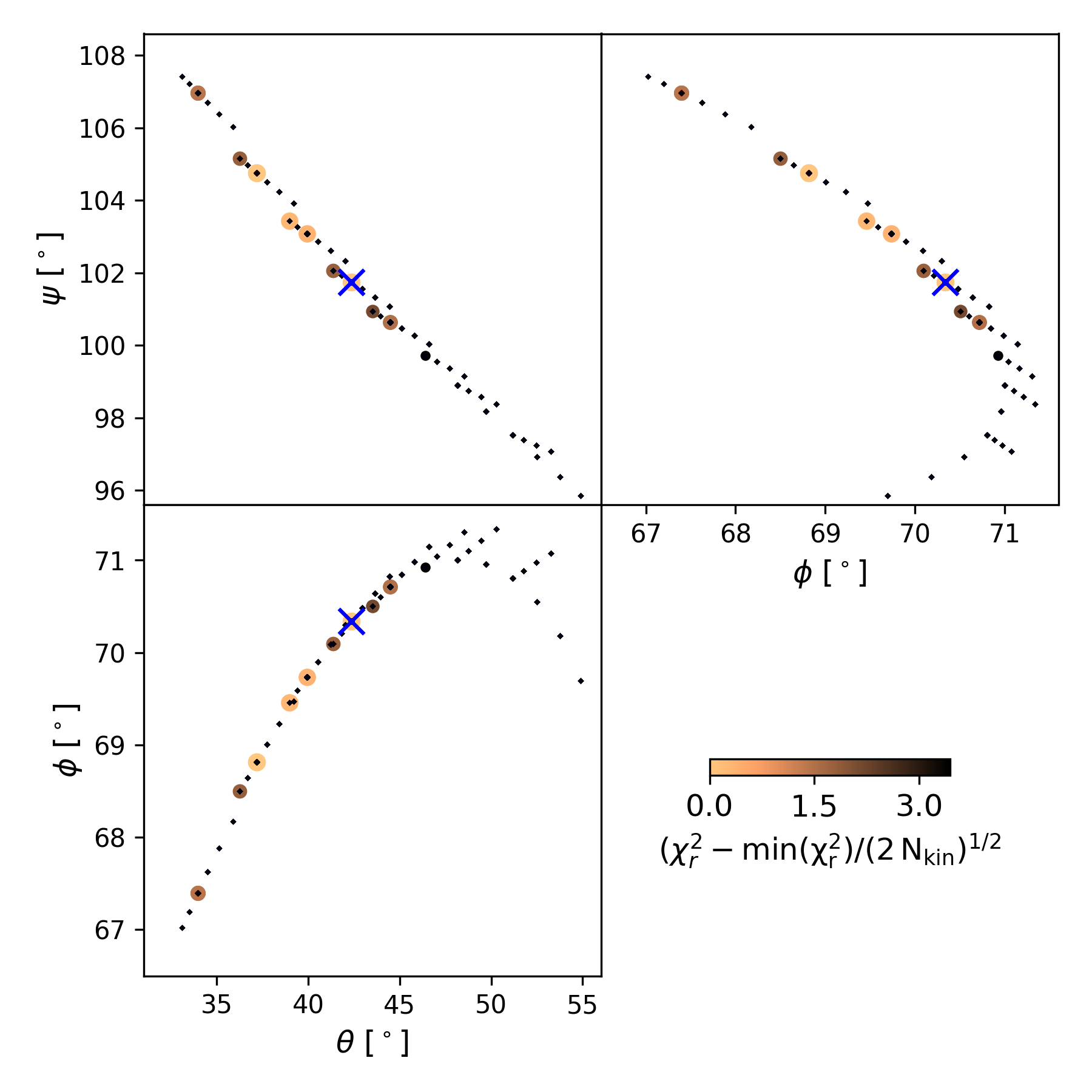}
\caption{Parameter grid of allowed viewing angles from our Schwarzschild modelling. The color and the symbol size give the normalized reduced $\chi^2$. The best-fit model is marked by the blue cross.}
\label{fig:angles}
\end{figure}

We show the grid of allowed viewing angles $\theta$, $\phi$ and $\psi$ in Fig. \ref{fig:angles}. Each dot corresponds to a single model and the best-fitting model is given by the blue cross. Due to the KDCs in the kinematics, the viewing angles are well constrained and thus only a limited range of deprojections from 2D to 3D are feasible.

\subsection{Properties of the extracted GCs}
We list coordinates, magnitudes, spectral S/N, radial velocities and - if possible - metallicities of the extracted GCs in Table \ref{tab:GC_properties}. The table includes all GCs with S/N $\geq$ 3. 

\begin{table*}
    \centering
    \begin{tabular}{c c c c c c c c}
         ID & RA & DEC & $g$ & $g - z$ & S/N & $v_\text{LOS}$ & [M/H] \\ 
            & (J2000) & (J2000) & (mag) & (mag) & (\AA$^{-1}$) & (km s$^{-1}$) &    (dex)  \\ \hline \hline
 FCC\,47-UCD1 & 03:26:32.92 & $-$35:42:56.09  & 20.8 & 0.75 & 20 & 1509.7 $\pm$ 4.5 & $-$1.12 $\pm$ 0.10 \\
1 & 03:26:31.96 & $-$35:42:50.59 & 24.8 & 0.88 & 8.4 & 1445.6 $\pm$ 8.0 & -- \\
2 & 03:26:32.34 & $-$35:42:44.42 & 22.6 & 0.95 & 5.5 & 1427.8 $\pm$ 18.2 & -- \\
3 & 03:26:32.32 & $-$35:42:44.15 & 25.1 & 1.41 & 6.4 & 1419.1 $\pm$ 16.2 & -- \\
4 & 03:26:31.87 & $-$35:42:45.26 & 24.5 & 1.19 & 4.9 & 1471.4 $\pm$ 16.7 & -- \\
5 & 03:26:32.00 & $-$35:42:43.71 & 23.4 & 1.26 & 7.1 & 1411.2 $\pm$ 8.5 & --  \\
6 & 03:26:31.82 & $-$35:42:44.99 & 23.8 & 1.01 & 5.0 & 1471.1 $\pm$ 8.7 & -- \\
15 & 03:26:31.51 & $-$35:43:03.59 & 23.3 & 0.39 & 4.0 & 1540.5 $\pm$ 22.5 & -- \\
17 & 03:26:31.46 & $-$35:43:04.42 & 22.7 & 1.14 & 7.4 & 1586.0 $\pm$ 27.1 & -- \\
18 & 03:26:33.39 & $-$35:43:01.53 & 21.5 & 1.03 & 15.0 & 1319.1 $\pm$ 3.6 & $-$0.84 $\pm$ 0.09 \\
19 & 03:26:32.38 & $-$35:43:08.69 & 23.5 & 0.87 & 3.0 & 1355.9 $\pm$ 44.5 & -- \\
23 & 03:26:32.49 & $-$35:43:14.45 & 22.6 & 1.30 & 6.6 & 1442.8 $\pm$ 5.8 & -- \\
26 & 03:26:30.39 & $-$35:43:10.66 & 23.2 & 1.11 & 3.6 & 1461.6 $\pm$ 13.9 & --\\
27 & 03:26:33.06 & $-$35:43:17.93 & 22.6 & 1.10 & 7.3 & 1365.1 $\pm$ 6.3 & -- \\
28 & 03:26:33.95 & $-$35:43:11.24 & 23.1 & 0.93 & 4.1 & 1465.2 $\pm$ 16.2 & --\\
29 & 03:26:29.88 & $-$35:43:04.71 & 22.8 & 1.27 & 6.1 & 1450.2 $\pm$ 8.3 & -- \\
30 & 03:26:30.26 & $-$35:43:15.72 & 23.1 & 1.01 & 3.6 & 1439.6 $\pm$ 16.1 & -- \\
32  & 03:26:31.31 & $-$35:43:27.91 & 23.3 &  0.94  & 3.5 & 1365.7 $\pm$ 40.6 & -- \\
33 & 03:26:34.32 & $-$35:43:20.06 & 22.7 & 0.96 & 6.2 & 1434.0 $\pm$ 14.3 & -- \\
34 & 03:26:30.03 & $-$35:43:21.50 & 21.5 & 0.87 &14.0 & 1442.7 $\pm$ 5.7 & $-$1.54 $\pm$ 0.11 \\
35 & 03:26:32.47 & $-$35:43:30.33 & 23.1 & 0.96 & 4.6 & 1490.2 $\pm$ 13.8 & --\\
36 & 03:26:29.27 & $-$35:43:19.04 & 22.8 & 0.87 & 4.8 & 1484.0 $\pm$ 17.7 & -- \\
38 & 03:26:29.83 & $-$35:43:32.14 & 21.4 & 0.95 & 16.0 & 1510.1 $\pm$ 3.9 & $-$1.19 $\pm$ 0.09 \\
40 & 03:26:33.10 & $-$35:43:43.80 & 22.4 & 1.29 & 10.0 & 1451.4 $\pm$ 5.0 & $-$0.22 $\pm$ 0.13 \\
42 & 03:26:30.82 & $-$35:43:47.26 & 21.8 & 1.06 & 12.7 & 1433.5 $\pm$ 4.0 & $-$0.95 $\pm$ 0.14\\
        \hline
    \end{tabular}
    \caption{ID, Coordinates, $g$-band magnitudes from \cite{Jordan2015} with typical uncertainties of $< 0.01$ mag, $g - z$ colour, S/N, LOS velocities and mean metallicities of the GCs in FCC\,47. We only show the GCs with S/N $\geq$ 3. The metallicities refer to our measurements with E-MILES SSP templates. The properties of FCC\,47-UCD1 are from \citet{Fahrion2019}.}
    \label{tab:GC_properties}
\end{table*}

\end{document}